\documentclass[aip,reprint]{revtex4-1}
\usepackage{hyperref}
\usepackage{graphics,graphicx,epsfig}
\usepackage{epsf,epstopdf,wrapfig}
\usepackage{amsmath,amssymb}
\usepackage{xcolor}
\usepackage{dsfont}

\usepackage{newfloat}
\usepackage{algorithm}
\usepackage{algpseudocode}
\usepackage{mathtools}

\draft 

\begin{document}


\title{Maximally predictive states: from partial observations to long timescales}

\author{Antonio C. Costa}
\email{antonio.costa@phys.ens.fr}
\altaffiliation{Current address: Laboratoire de Physique de l’Ecole normale supérieure, ENS, Université PSL, CNRS, Sorbonne Université, Université Paris Citéß, F-75005 Paris, France}
\affiliation{Department of Physics and Astronomy, Vrije Universiteit Amsterdam, 1081HV  Amsterdam, The Netherlands}
\author{Tosif Ahamed}
\affiliation{Lunenfeld-Tanenbaum Research Institute, Mount Sinai Hospital, Toronto, Canada}
\author{David Jordan}
\affiliation{Wellcome/CRUK Gurdon Institute, University of Cambridge, United Kingdom}
\author{Greg J. Stephens}
\affiliation{Department of Physics and Astronomy, Vrije Universiteit Amsterdam, 1081HV  Amsterdam, The Netherlands}
\affiliation{Biological Physics Theory Unit, OIST Graduate University, Okinawa 904-0495, Japan}

\date{\today}

\begin{abstract}
Isolating slower dynamics from fast fluctuations has proven remarkably powerful, but how do we proceed from partial observations of dynamical systems for which we lack underlying equations? Here, we construct maximally-predictive states by concatenating measurements in time, partitioning the resulting sequences using maximum entropy, and choosing the sequence length to maximize short-time predictive information. Transitions between these states yield a simple approximation of the transfer operator, which we use to reveal timescale separation and long-lived collective modes through the operator spectrum. Applicable to both deterministic and stochastic processes, we illustrate our approach through partial observations of the Lorenz system and the stochastic dynamics of a particle in a double-well potential. We use our transfer operator approach to provide a new estimator of the Kolmogorov-Sinai entropy, which we demonstrate in discrete and continuous-time systems, as well as the movement behavior of the nematode worm {\em C. elegans}.
\end{abstract}


\maketitle 

\begin{quotation}
Complex structure often arises from a limited set of simpler elements, such as novels from letters or proteins from amino acids. In musical composition, for example, we experience such construction in time; sounds and silences combine to form motifs; motifs form passages which in turn form movements. But how can we generally identify variables which distinguish structures across timescales, especially under typical conditions of nonlinear dynamics which are measured only incompletely? Here, we introduce a framework for finding effective coarse-grained variables by leveraging the transfer operator evolution of ensembles. Just as a musical piece transitions from one movement to the next, the ensemble dynamics consists of transitions between local collections of states. We use operator dynamics to guide the construction of maximally predictive states from incomplete measurements, a reconstruction and partitioning of the underlying state space. We then identify timescale separated, coarse-grained variables through operator eigenvalue decomposition. The generality of our approach provides an opportunity for insights on long term dynamics within a wide variety of complex systems.
\end{quotation}

\section{Introduction}

The constituents of the natural world combine to form a dizzying variety of emergent structures across a wide range of scales.  Sometimes these structures are obvious without referencing their underlying components; we need not, famously, link bulldozers to quarks \cite{Goldenfeld1999}. However, many are more ambiguous, especially in complex systems.  In these cases connecting across scales can provide greater insight than from either scale alone. 

Statistical physics provides a number of successful examples in which the principled integration of fine-scale degrees of freedom gives rise to coarse-grained theories that successfully capture and predict larger scale structure. However powerful, such approaches generally require either a deep understanding of the underlying dynamics, symmetries and conservation laws or an appropriate parameterization for methods like the renormalization group \cite{Goldenfeld1992}. We aim to emulate this success but working directly from data in systems which permit neither detailed equations of motion nor obvious scale separations.  How can we find effective discretizations and coarse-grainings in such partially observed complex dynamical systems? 

In the absence of theory, measurements typically constitute only a partial observation of the dynamics: one cannot know \emph{a priori} the relevant degrees of freedom to measure. This poses fundamental challenges to predictive modeling. Indeed, successful forecasting of time series data typically requires history-dependent terms, either explicitly through autoregressive models \cite{Box1976} or more implicitly through recurrent neural networks or hidden Markov models \cite{Uribarri2022,Bhat2022}, for example. These methods are justified mathematically through delay-embedding theorems developed by Takens and others \cite{Takens1981,Sauer1991,Stark1999,Stark2003,Sugihara1990}, which state that given a generic measurement of a dynamical system, it is possible to reconstruct the state space by concatenating enough time delays of the measurement data. Building on these results, we search for principled coarse-grainings of the dynamics from partial observations. 

We trade individual trajectories for ensembles, and seek long-lived dynamics in the patterns of stochastic state-space transitions.  Formally, we apply the machinery of \emph{transfer operators} \cite{Koopman1931,Koopman1931,Mezic2004,Mezic2005,Bollt2013} to simultaneously reconstruct the dynamics and learn effective coarse-grained descriptions on long timescales. In this framework, nonlinear dynamics are captured through linear yet infinite-dimensional differential operators; the Fokker-Planck equation is a simple example. We focus on the discrete-time evolution of the probability density, which is governed by Perron-Frobenius (PF) operators \cite{Gaspard1998,Givon2004,Pavliotis2014,Lasota1994} that ``transfer'' state-space densities forward in time. 

Transfer operators are invariant to smooth transformations of the state-space coordinates \cite{Berry2013,Arbabi2017,Das2019,Giannakis2019}. This makes them especially convenient when working directly with incomplete time series measurements, from which it is generally possible to obtain a topologically equivalent state-space reconstruction through a delay embedding \cite{Takens1981,Stark1999,Sauer1991,Deyle2011,Yair2017,Gilani2021}. 
In addition to Fokker-Planck, transfer operators appear in other white noise processes and even deterministic dynamics \cite{Lasota1994,Bollt2011, Pavliotis2014}, such as the Liouville operator of classical Hamiltonian dynamics. We leverage this universality to study stochastic and deterministic systems through a single framework.

The operator approach affords an additional important advantage; the timescales of the system are naturally ordered by the eigenvalue spectrum, even when this ordering is hidden or subtle in the original observations.  Correspondingly, the eigenfunctions capture emergent coarse-grained dynamics such as transitions between metastable states.  In the context of chemical kinetics, for example, operator eigenfunctions have been shown to provide effective reaction coordinates or order parameters
\cite{Perez-Hernandez2013,Chodera2014,Bittracher2018a}. 

Here, from partial observations of a complex dynamical system, we simultaneously reconstruct a maximally Markovian state space through delay embedding and partition the resulting reconstruction to obtain a finite approximation of the PF operator as a Markov transition matrix \cite{Lasota1994,Bollt2013}. We then use the eigendecomposition of the transition matrix to connect fine-scale and coarse-grained descriptions. In Sec. \ref{sec:MaxPred}, we introduce “maximally predictive states”, a partitioning constructed by optimizing both the number of time delays and the spatial discretization scales given the constraints imposed by the data. In Sec. \ref{sec:coarse-graining}, we coarse-grain the dynamics by using the eigenspectrum of the transition matrix to maximize the dynamical coherence of groups of fine-scale states. Finally, in Sec. \ref{sec:KS_entropy} we use our transfer operator dynamics to provide a novel estimator of the Kolmogorov-Sinai (KS) entropy rate.

\begin{figure*}
\begin{center}
\includegraphics[scale=1.]{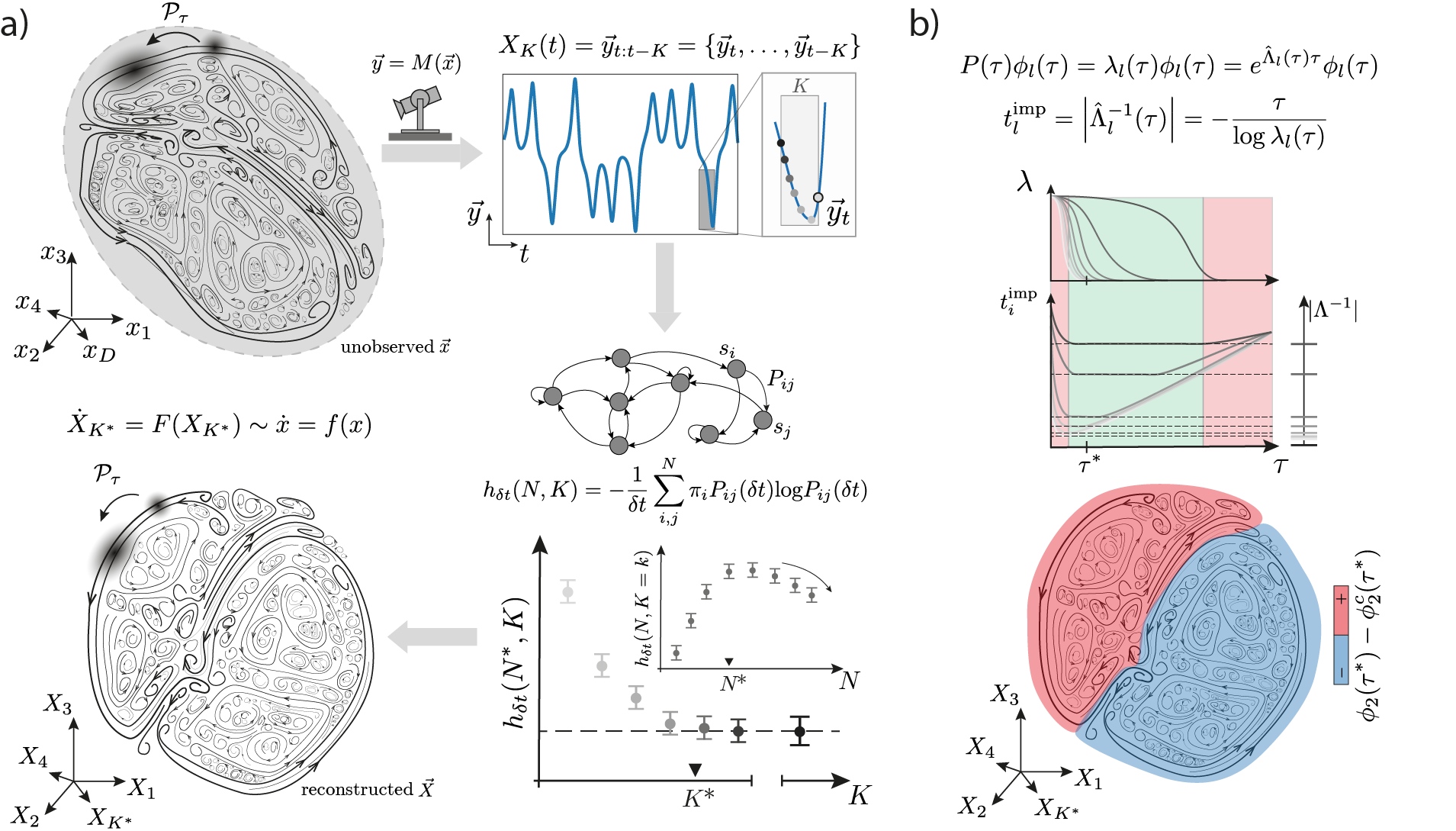}
\caption{{\bf Maximally predictive states for coarse-graining through ensemble dynamics.}
(a) Given partial observations $\vec{y} = M(\vec{x})$ of an unknown dynamical system $\dot{x} = f(x)$, we reconstruct the state space by concatenating measurements in time to maximize predictability. For each delay $K$ we construct an embedding $X_K$ (top right) and partition this space to approximate the Perron-Frobenius operator as a Markov chain $P_{ij}$ (middle). We choose the number of partitions $N^*$ to maximize the short-time entropy rate $h_{\delta t}(N,K)$, up to finite-size effects and find the $K^*$ that minimizes the unpredictability of the reconstructed transfer operator dynamics as measured through $h_{\delta t}(N^*,K)$ (bottom right). An appropriate embedding results in a reconstructed state that preserves the properties of $\mathcal{P}_\tau$ (bottom left).
(b-top) For an appropriate waiting time $\tau^*$, The eigenspectrum of $P_{ij}(\tau^*)$ reveals the timescales of the system. The inferred relaxation times of the underlying dynamics, $t_\text{imp}$, implicitly depend on the choice of $\tau$. For $\tau\rightarrow 0$, $t_\text{imp}$ exhibit a short transient; $P_{ij}$ is close to the identity matrix and the eigenvalues are nearly degenerate about 1 (left red region). For $\tau\gg\tau^*$ the eigenvalues are approximately constant as $P_{ij}$ is determined by fluctuations about the invariant distribution and the relaxation times grow linearly (right red region). We choose $\tau^*$ such that the long-time dynamics are approximately Markovian and the spectral gap is the largest (green shaded region). 
(b-bottom).  We use $P_{ij}(\tau^*)$ to identify coherent sets by finding the partition along the slowest eigenvector $\phi_2(\tau^*)$ that maximizes the overall coherence of the sets.
}
\label{fig_1}
\end{center}
\end{figure*}

\section{\label{sec:MaxPred} Maximally predictive state-space reconstruction}

We consider the general setting of the time evolution of a system's state $\vec{x}$ through a set of differential equations $\dot{\vec{x}} = f(\vec{x})$, where $f$ encompasses both deterministic and noisy influences. Measurement data typically results from a  generic nonlinear, possibly noisy, observation function \cite{Aeyeles1981,Takens1998} $M$  that maps the unobserved state $\vec{x}(t) \in \mathbb{R}^D$, $t\in \mathbb{R}^+_0$ into a discrete time series $\vec{y}(t) = M(\vec{x}(t)) \in \mathbb{R}^d$, $t = \{0,\delta t,\ldots,T\}$, where generally $d< D$. This means that we typically only measure a subset of the variables that define the state $\vec{x}$, with finite precision. The unobserved degrees of freedom induce history-dependence on $\vec{y}$(t), which not only depends on $\vec{y}(t-1)$ but also on past observations. Formally, the measurement function $M$ imposes in a Mori-Zwanzig projection of the dynamics \cite{Mori1965,Zwanzig1973} such that
\begin{align*}
    \vec{y}_{t+1} = f_0 (\vec{y}_t) + \sum_{k=1}^{K^*}f_k(\vec{y}_{t-k}) + \eta_t,
\end{align*}
where $f_0$ is a Markovian term, $f_k$ capture non-Markovian effects over a timescale $K^*$ and $\eta_t$ are fluctuations orthogonal to the projection $M$. This means that neglecting the history dependence imposed by this projection results in a systematic source of error \cite{Chorin2002,Rupe2022,Gilani2021}, which we must resolve in order to build an accurate predictive model of the dynamics.

Instead of modeling $f_k$ explicitly, in our approach we search for a representation in which the dynamics is maximally Markovian. We concatenate $K-1$ time delays of the measurement time series, yielding a candidate state space $X_K(t) \in \mathbb{R}^{d\times K}$, Fig.~\ref{fig_1}(a). Including past information (larger $K$) decreases the uncertainty of the immediate future, thus increasing the predictability of the state space. We search for the number of delays $K^*$ that maximize this predictability, as quantified through the short-time entropy rate \cite{Packard1980}.

We bin $X = \{X_K(0),X_K(\delta t),\ldots,X_K(T)\}$ into a discrete partitioned space with $N$ Voronoi cells, $s_X =\{s_i(0),s_j(\delta t),\ldots,s_k(T)\}$ and estimate the short-time entropy rate\footnote{Not to be confused with the KS entropy \cite{Kolmogorov1958}, which poses a fundamental bound to the predictability of a dynamical system. We note that our inferred Markov chain is \emph{not} a complete model of the dynamics, but only an approximate description on time scales and length scales beyond the transition time $\tau$ and the length scale imposed by the partitioning $\epsilon \sim 1/N$. In principle, there exist \emph{generating partitions} that exactly preserve the continuum dynamics, but these are generally challenging to find except for simple one or two-dimensional discrete maps \cite{Rokhlin1967,Bollt2013}. We return to this point in Sec. \ref{sec:KS_entropy}.} as

\begin{equation}\label{eq:h_p}
   h_{\delta t}(N,K) =  -\frac{1}{\delta t} \sum_{i,j}^N \pi_{i} P_{ij}(\delta t) \text{log}P_{ij} (\delta t)
\end{equation}
where $P_{ij}(\delta t)  = p\left(s_j(t+\delta t)|s_i(t)\right)$ is a row-stochastic Markov chain and $\pi$ is the stationary distribution of $P$, Fig.~\ref{fig_1}(a). We note that $P$ is an approximation of the PF transfer operator that evolves densities on the reconstructed state space (Methods). 

The behavior of $h_{\delta t}(N,K)$ with $N$, or equivalently, with a typical length scale $\epsilon \sim 1/N$, is a non-decreasing function that is indicative of different classes of dynamics \cite{Gaspard1993}. For example, stochastic processes possess information on all length scales, and so $h_{\delta t}(N\rightarrow\infty,K)=\infty$, whereas the fractal nature of deterministic chaotic systems yields a typical length scale below which the entropy stops changing: $h_{\delta t}(N^* = 1/\epsilon^*,K)=h_{\delta t}(\infty,K)$. In practice however, finite-size effects yield an underestimation of the entropy rate for large $N$. With the aim of preserving as much information as possible given the data and our partitioning scheme, we set the number of partitions as the largest $N$ after which the entropy rate stops increasing, Fig.~\ref{fig_1}(a). We thus maximize the entropy with respect to the number of partitions, obtaining $ h_{\delta t}(K) = \max_N  h_{\delta t}(N,K)$. 

For sufficiently short $K\delta t$, $h_{\delta t}(K)$ is a monotonically non-increasing function of $K$, and so, we add time delays until we find $K^*$ such that $\delta h = h_{\delta t}(K^*+1)-h_{\delta t}(K^*) \approx 0$, capturing the amount of memory sitting in the incomplete measurements \cite{Lu2016,Kantz2000,Gilani2021} and obtaining a maximally predictive description of the dynamics. We note that $\delta h$ has been previously used to define measures of forecasting complexity in dynamical systems \cite{Grassberger1986} and is the amount of information that has to be kept in the $K-1$ time delays for an accurate forecast of the next time step.  Notably, $K^*$ is related to the intrinsic dimension of the state space $D$: given a generic measurement function \cite{Aeyeles1981,Takens1998} $M$, delay embedding theorems \cite{Takens1981,Sauer1991,Stark1999,Stark2003,Sugihara1990} guarantee a Markovian reconstruction when the number of time delays is larger than twice the dimension of the underlying state-space, $K>2D$.

In the following section, we illustrate our approach to build maximally-predictive states with applications in stochastic equilibrium dynamics of a particle in a double-well potential and the dissipative chaotic dynamics of the Lorenz system.

\subsection*{Illustrative applications}

\begin{figure}[ht!]
\begin{center}
\includegraphics[scale=1.05]{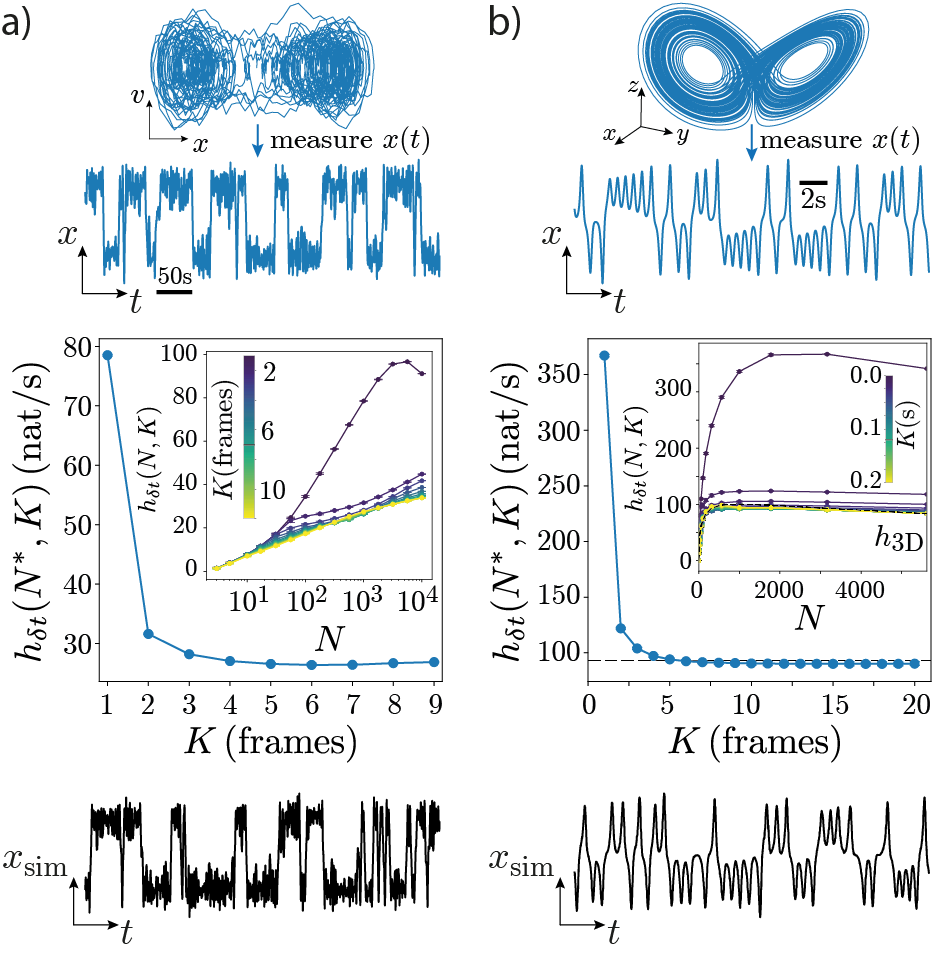}
\caption{{\bf Maximally predictive state-space reconstruction in stochastic and deterministic examples.}
(a-top) We simulate the Langevin dynamics of an underdamped particle in a double-well potential Eq.\,(\ref{eq:DW_phspace}) and measure only $x(t)$ sampled at $\delta t = 0.05\,\text{s}$; The example time series has $\beta^{-1}=0.5\,\text{J}$.
(a-middle) Short-time entropy rate $h_{\delta t}(N^*,K)$, with $N^*=10^3\,\text{partitions}$ chosen to avoid finite-size effects (inset). Underdamped dynamics are evident as the short-time entropy rate with $K=1$ is much larger than with additional delays. We choose $K^*=7\,\text{frames}$. 
(a-bottom) Simulation of the Markov chain inferred from partial observations of the double-well dynamics.
(b-top) We simulate the Lorenz system Eq.\,(\ref{eq:Lorenz}) in the standard chaotic regime, and measure only $x(t)$, sampled at $\delta t = 0.01\,\text{s}$.
(b-middle) Short-time entropy rate $h_{\delta t}(N^*,K)$, with $N^*=10^{3.5}=3162\,\text{partitions}$, chosen so as to avoid finite-size effects (inset). After $K\sim 10\,\text{frames}$ we reach a maximally predictive state, which converges to estimates from the full $(x,y,z)$ state space (black dashed line); we use $K^* = 12\,\text{frames} = 0.12\,\text{s}$.
(b-bottom) Simulation of the Markov chain inferred from partial observations of the Lorenz system.
Error bars ($<$ marker size) are $95\%$ confidence intervals bootstrapped over non-overlapping 2000\,s segments in (a) and  $50\,000\,\text{s}$ segments in (b).
}
\label{fig_2}
\end{center}
\end{figure}

Consider the underdamped Langevin dynamics of a particle in a double-well potential at thermal equilibrium, Fig.\,\ref{fig_2}(a-top),

\begin{align}\label{eq:DW_phspace}
    \begin{cases}
    dx_t  =   v_t dt  \\
    dv_t  =   -v_t dt -4x_t(x_t^2-1)dt+\sqrt{2 \beta^{-1}}dW_t, 
    \end{cases}
\end{align}

\noindent where $\beta^{-1} = k_B T$, $k_B$ and $T$ are the Boltzmann constant and temperature, respectively, and $dW_t$ is the Wiener process capturing the much faster random collisions with the heat bath. We sample at $\delta t = 0.05\,\text{s}$ for $T=10^6\,\text{s}$, with temperatures ranging between $\beta^{-1} = 0.5\,\text{J}$ and $\beta^{-1} = 2.5\,\text{J}$ (Appendix \ref{sec:Simulations}): an example trajectory for $\beta^{-1} = 0.5\,\text{J}$ is shown in Fig.\,\ref{fig_2}(a-top). We emulate a real data example by sampling only $x$ at discrete time steps, from which we seek to reconstruct the state space by including time delays. For simplicity, we here take a linear measurement function with noise coming from the finite precision of the simulation protocol. Nonetheless, our results generalize to more complex functions and added measurement noise, as previously discussed in, e.g., Refs. \onlinecite{Casdagli1991,Gibson1992}. We concatenate the $x$ measurement with $K$ delays, partition the resulting state with $N$ partitions and study the behavior of the short-time entropy rate as a function $K$ and $N$, Fig.~\ref{fig_2}(a-middle), Fig.~S1(a). The stochastic nature of the dynamics yields $\lim_{N\rightarrow\infty} h_{\delta t}(N,K) = \infty$, up to finite-sampling effects which produce an underestimation of the entropy\cite{Cohen1985,Gaspard1993} for large $N$, a drop we see most visibly for $K=1\,\text{frame}$. At intermediate $N$, where estimation effects are negligible, there is an abrupt change in the entropy rate from $K=1$ to $K>2\,\text{frames}=0.1\,\text{s}$, indicative of the finite memory sitting in the measurement time series \cite{Gloter2006,Pavliotis2007,Ferretti2020,Bruckner2020} from the missing degree of freedom\footnote{Notably, the dynamics is not fully memoryless with $K=2$ delays, as one would naively expect, a result due to the Euler-scheme update which induces memory effects on the short-time propagator $P(x_{k+1}|x_k,x_{k-1})$ \cite{Gloter2006,Pavliotis2007,Ferretti2020,Bruckner2020}}. The behavior of the entropy with $K$ and $N$ is qualitatively conserved across the range of temperatures studied here, Fig.~S2(a), and we choose $K^* = 7\,\text{frames}=0.35\,\text{s}$ for subsequent analysis. Our reconstructed state recovers the equipartition theorem, Fig.\,S2(b), showing that momentum information is accurately captured by the reconstructed state. In addition, our effective model with $K^*$ produces realistic simulations of the position dynamics, as illustrated in Fig.\,\ref{fig_2}(a-bottom), Fig.\,S3(a).

Consider next the chaotic dynamics of the Lorenz system \cite{Lorenz1963},
\begin{align}\label{eq:Lorenz}
    \begin{cases}
    \dot{x} = & \sigma(y-x)  \\
    \dot{y} = & x(\rho-z) - y\\
    \dot{z} = & xy - \beta z
    \end{cases},
\end{align}
\noindent with $\sigma = 10$, $\rho=28$ and $\beta = 8/3$, Fig.\,\ref{fig_2}(b-top). Unlike the double-well dynamics, the fractal structure of the Lorenz system does not technically satisfy the requirements of our modeling approach, since our partition scheme (into disjoint non-overlapping sets) assumes a smooth invariant measure rather than a fractal set. Nonetheless, we will show how we can still use the same exact framework to accurately reconstruct the state space of the Lorenz system and recover effective coarse-grained states from partial observations.

We measure only $x$ at $\delta t = 0.01\,\text{s}$ for $T = 5\times 10^5\,\text{s}$ (Appendix \ref{sec:Simulations}) and reconstruct the state space as before: concatenating $K$ time delays and studying the behavior of the short-time entropy rate with $N$ and $K$, Figs.\,\ref{fig_2}(b-middle). As expected for deterministic chaos \cite{Gaspard1993}, we observe that the short-time entropy rate reaches an asymptotic value for increasing $N$, $h_{\delta t}(N>N^*,K)\approx h_{\delta t}(N^*,K)$. As a function of $K$, the short-time entropy rate exhibits a sharp decrease, becoming constant after $K\gtrsim10\,\text{frames}$, Fig.\,S1(b), indicating that we have reached a maximally predictive state. We use $K^*=12\,\text{frames}=0.12\,\text{s}$ to reconstruct the state space. Importantly, we note that our partition of the state space results in an over-estimation of the Kolmogorov-Sinai entropy, only attainable in the limit $\delta t\rightarrow 0$ and $\epsilon \sim 1/N \rightarrow 0$, a point we return to in Sec. \ref{sec:KS_entropy}. Nonetheless, the state space reconstructed with our choice of $K^*$ is equivalent to the underlying state of the system: the short-time entropy rate computed from the underlying $(x,y,z)$ state space across $N$ matches that of the state reconstructed with $K^*$ time delays (dashed black lines in Fig.~\ref{fig_2}(b-middle)). In addition, the inferred maximally predictive states and the resulting Markov approximation of the dynamics allow us to generate realistic simulations of the $x$ time series, as illustrated in Fig.\,\ref{fig_2}(b-bottom), Fig.\,S3(b). Finally, we also note that the reconstructed transfer operator dynamics yields accurate estimates of the information dimension \cite{Farmer1983}, Fig.\,S4.

\section{\label{sec:coarse-graining} Extracting coarse-grained dynamics}

In the state-space reconstruction we encode ensemble dynamics in a matrix $P_{ij}$ constructed by counting transitions between cells $s_i$ and $s_j$ in a transition time $\tau$: formally this is a discrete-space approximation to the PF operator $\mathcal{P}_\tau$ \cite{Bollt2013}, Fig.~\ref{fig_1}(a-right). This object plays a central role in our analysis: it not only guides the state-space reconstruction but also provides means for a principled coarse-graining of the state space through its eigenvalue spectrum. The  eigenvalues of $P_{ij}$ directly reveal timescale separation and the corresponding eigenvectors can be used to identify regions of state space where the system lingers, these are ``macroscopic'' metastable states \cite{Bollt2013,Schutte2001}. The structure of these regions and the kinetics between them offer a principled coarse-graining of the original dynamics. 

\subsection*{Choosing a transition time}

While the instantaneous ensemble dynamics corresponding to $\dot{x} = f(x)$ is given by $\dot{\rho} = \mathcal{L}\rho$, the finite-time transfer operator $\rho_{t+\tau} = \mathcal{P}_\tau\rho_t$ is more immediately available when working with discrete-time measurements. The operators $\mathcal{L}$ and $\mathcal{P}_\tau$ share the same set of eigenfunctions, while the eigenvalues $\lambda_i^*$ of $\mathcal{P}_\tau$ are exponential functions of eigenvalues $\Lambda^*_i$ of $\mathcal{L}$, $\lambda^*_{i} = e^{\Lambda^*_i \tau}$. When the estimated eigenvalues $\lambda_i$ are real, the relaxation of the inferred density dynamics is characterized by the implied relaxation times corresponding to each eigenvalue,

\begin{align}\label{eq:t_imp}
t^\text{imp}_i(\tau)=\left|\hat{\Lambda}_i^{-1} \right| (\tau) = \frac{-\tau}{\log \lambda_{i}(\tau)}.
\end{align}

$\mathcal{P}_\tau$ lives in an infinite-dimensional functional space, which we discretize using $N$ basis functions: in our partitioned space, the basis functions are characteristic functions and the measure is piecewise constant (Appendix \ref{sec:Methods}). The truncation at finite $N$ erases fine-scale information within the partition, and so the ability of the transfer operator approximation $P_{ij}$ to capture the large-scale dynamics depends on the transition time $\tau$, a parameter that we vary. For the state-space reconstruction described in the Sec.\,\ref{sec:MaxPred}, we chose $\tau$ as the sampling time $\tau=\delta t$ in order to maximize short-time predictability. However, to accurately capture longer-time dynamics and metastable states, we vary $\tau$ and study changes in the inferred spectrum, Fig.\,\ref{fig_1}(b).  

For $\tau\rightarrow 0$, $P_{ij}$ is close to an identity matrix (the system has little time to leave its current partition) and all eigenvalues are nearly degenerate and close to unity, Fig.~\ref{fig_1}(b). For $\tau$ much longer than the mixing time of the system, the eigenvalues start collapsing and in the limit of $\tau\rightarrow\infty\Rightarrow P_{ij}(\tau) \sim \pi_j$: the probability of two subsequent states separated by such $\tau$ becomes independent, the eigenvalues of $P_{ij}(\tau)$ stop changing and $t_\text{imp}$ grows linearly with $\tau$ for all eigenfunctions. In this regime, the transition probability matrix contains noisy copies of the invariant density akin to a shuffle of the symbolic sequence. This yields an effective noise floor, which is observed earlier (shorter $\tau$) for faster decaying eigenfunctions. For intermediate $\tau$ we find a spectral gap, indicating that the fast dynamics have relaxed and we can isolate the slow dynamics. In addition, the Markovian nature of the dynamics implies that the inferred relaxation times reach a constant value that matches the underlying long timescale $\Lambda_2^{-1}$ of the system  \footnote{We note that this is not a sufficient condition for Markovianity, as also the eigenvectors need to be constant with $\tau$.}: the Chapman-Kolmogorov equation (see e.g. Ref. \onlinecite{Papoulis1984}) is verified $\mathcal{P}_{n\tau}\rho = \mathcal{P}^n_\tau\rho$, and thus $t^\text{imp}_i(n\tau)=\frac{-n\tau}{\log (\lambda_{i}(\tau)^n)}=t^\text{imp}_i( \tau)$. We choose $\tau^*$ after the initial transient, searching for a nearly constant $\hat{\Lambda}_2$ and a large spectral gap (green shaded region in Fig.~\ref{fig_1}(b)). 

\begin{figure*}
\begin{center}
\includegraphics[scale=1]{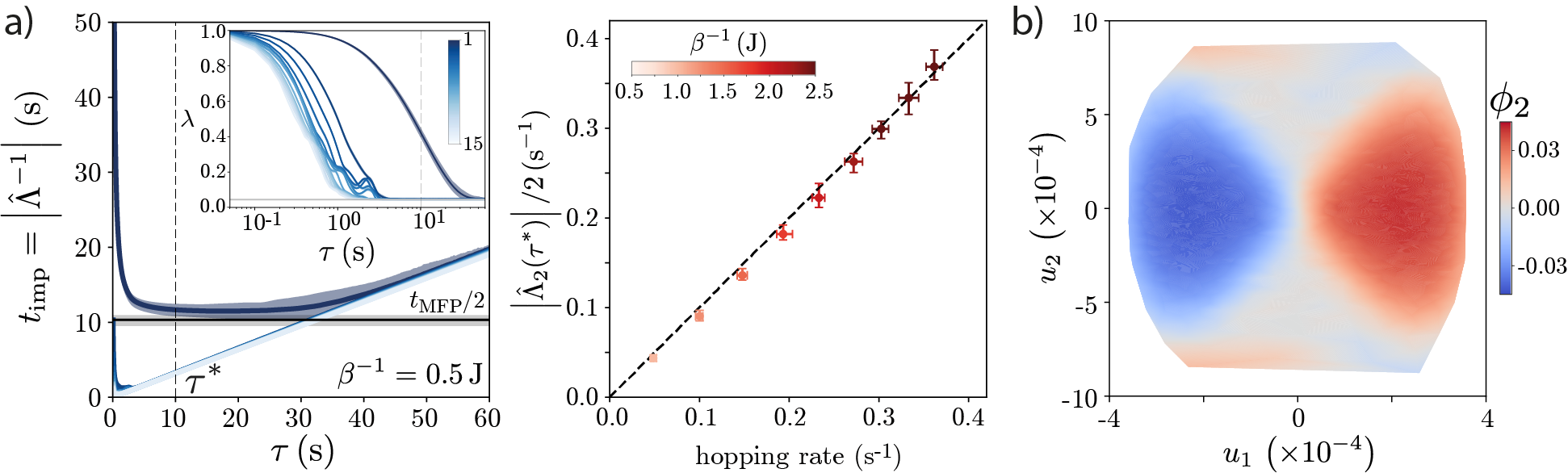}
\caption{{\bf Transfer operator dynamics and coherent sets in double-well, underdamped Langevin dynamics.}
(a-left) Implied timescales $t_\text{imp}$ vs.~transition time $\tau$ for the reconstructed dynamics of a particle in a double-well potential. We plot the eigenvalues $\lambda(\tau)$ of the (reversibilized) transfer operator ${P_r}(\tau)$ (inset) and corresponding implied timescale $t_\text{imp}(\tau)$  for $\beta^{-1} = k_B T = 0.5\,\text{J}$. Markov dynamics in the first non-trivial eigenfunction $\phi_2$ are evident when $t_\text{imp}$ is constant (dark blue). The short-$\tau$ transient results from a nearly diagonal transition matrix reflected in the accumulation of eigenvalues near 1. For large $\tau$, the eigenvalues stop changing as the transition probabilities reach the noise floor (gray horizontal line in inset). The longest relaxation time is approximately constant after $\tau\sim 5\,\text{s}$, converging to half the mean first passage time $t_\text{MFP}/2$ (black horizontal line).
(a-right) $|\hat{\Lambda}_2(\tau^*)|$ predicts the hopping rate across temperatures. We choose $\tau^*$ separately for each temperature to reflect the different dynamics (see Fig.\,S5(b)). Error bars are 95\% confidence intervals bootstrapped over $50\,000\,\text{s}$ non-overlapping segments.
(b) Contour plot of the first non-trivial eigenvector $\phi_2$, projected onto the two SVD modes $(u_1,u_2)$ with largest singular values of the state-space delay reconstruction, here for $\beta^{-1}=0.5\,\text{J}$, Fig.\,S2(c). The sign of $\phi_2$ effectively splits the reconstructed state space into the two wells. 
}
\label{fig_3_DW}
\end{center}
\end{figure*}

\subsection*{Identifying metastable states}
Within our reconstructed ensemble evolution, coarse-grained dynamics can be identified between state-space regions where the system becomes temporarily trapped, these are metastable states or almost-invariant sets \cite{Dellnitz1999,Deuflhard2000}. We search for collections of states that move coherently with the flow: their relative future states belong to the same macroscopic set. Since the slowest decay to the stationary distribution is captured by the first non-trivial eigenfunction of the transfer operator, we search for the subdivision along this eigenfunction that maximizes the coherence of the resulting macroscopic sets, a previously introduced heuristic \cite{Dellnitz1999}. 

A set $S$ is coherent when the system is more likely to remain within the set than it is to leave it within a time $\tau$. We quantify this intuition by measuring the overlap between a set of states and their time evolution through 
$\chi_{\mu,\tau}(S) = \frac{\mu(S\cap \Phi_{-\tau}S)}{\mu(S)} = \frac{\mu(\Phi_\tau S\cap S)}{\mu(S)}$,
where $\mu$ is the invariant measure preserved by the invertible flow $\Phi_\tau$. Given an inferred transfer operator $P_{ij}(\tau)$ and its associated stationary eigenvector $\pi$, we can immediately compute $\chi$ (Appendix \ref{sec:Methods}),
\begin{equation}\label{eq:coherence_practice}
    \chi_{\pi,\tau}(S) = \frac{\sum_{i,j\in S}\pi_i P_{ij}(\tau)}{\sum_{i \in S} \pi_i}.
\end{equation}

To identify optimally coherent metastable states through spectral analysis, we define a time-symmetric (reversibilized) transfer operator, $\mathcal{P}_r \equiv (\mathcal{P}+\mathcal{P}^\dagger)/2$,
where $\mathcal{P}^\dagger$ is the dual operator to $\mathcal{P}$, pulling the dynamics backward in time; see Appendix \ref{sec:Methods} for the discrete numerical approximation, $P_r$. While transfer operators describing ensemble dynamics are not generally symmetric \cite{Dellnitz1999,Froyland2003}, the definition of coherence is invariant under time reversal: since the measure is time invariant, it does not matter in which direction we look for mass loss from a set \cite{Froyland2005,Froyland2014}. Besides the invariance property, an important benefit of working with $P_r$ is that its second eigenvector $\phi_2$ provides an \emph{optimal} subdivision of the state space into almost-invariant sets, as shown in Ref.~\onlinecite{Froyland2005}. Given $\tau^*$ and the corresponding  $P_r(\tau^*)$, we identify metastable sets by choosing a subdivision along $\phi_2$ that maximizes an overall measure of coherence,
\begin{align}
\chi(\phi_2^c) = \text{min}\{\chi_{\pi,\tau^*}(S^+(\phi_2^c)),\chi_{\pi,\tau^*}(S^-(\phi_2^c))\},
\label{Eq:coherence_min}
\end{align}
where $\{S^+(\phi_2^c),S^-(\phi_2^c)\}$ result from a partition at $\phi_2^c$: the optimal almost-invariant sets are identified with respect to the sign of $\phi_2 - \phi_2^c$, Fig.\,\ref{fig_1}(b-bottom). In the following section, we illustrate the process of extracting coarse-grained dynamics from incomplete measurements.

\begin{figure}[ht!]
\begin{center}
\includegraphics[scale=1]{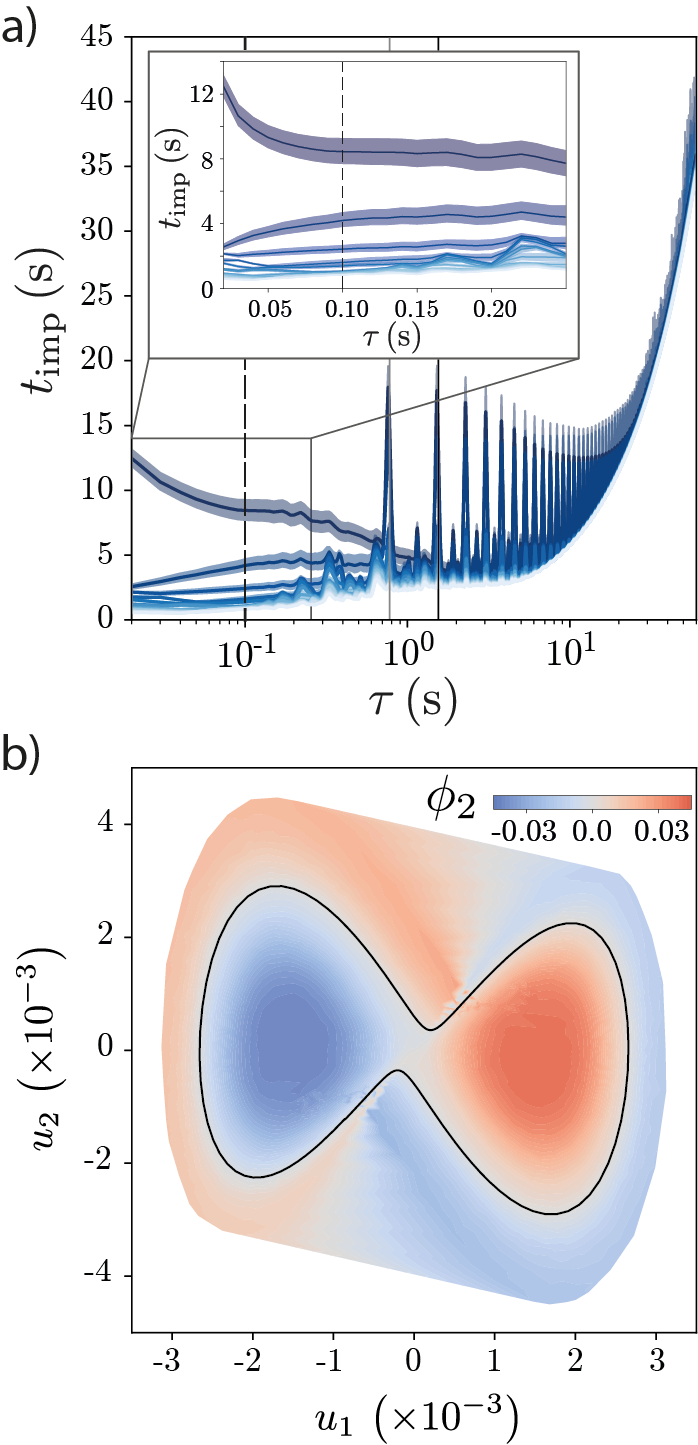}
\caption{{\bf Transfer operator dynamics and coherent sets in the Lorenz system.}
(a) Implied relaxation times vs.~$\tau$ for the reconstructed dynamics of the Lorenz system. There is an initial short transient before the longest relaxation time settles into a near constant value (inset). For much larger $\tau$, the quasi-periodicity of the Lorenz attractor results in an aliasing effect: the transfer operator becomes again nearly identity at integer multiples of $T_\text{UPO}/2$ (gray solid line, see Appendix \ref{sec:Methods}). Nonetheless, for very large $\tau$ the inferred transfer operator is composed of near copies of the stationary distribution, resulting in the collapse of the eigenvalues. We choose $\tau^* = 0.1\,\text{s}$. Error estimates are 95\% confidence intervals bootstrapped over 2000\,s non-overlapping segments. 
(b) Contour plot of the first non-trivial eigenvector $\phi_2$, projected onto the two SVD modes $(u_1,u_2)$ with largest singular values of the state space delay reconstruction. The sign of $\phi_2$ divides the state space into its almost-invariant sets, which are partially split along the shortest unstable periodic orbit \cite{Froyland2009}.
}
\label{fig_3_Lorenz}
\end{center}
\end{figure}

\subsection*{Illustrative applications}

We return to the equilibrium dynamics of a particle in a double-well potential, Eq.\,\ref{eq:DW_phspace}, Fig.\,\ref{fig_2}(a). Using the maximally predictive reconstructed state space, we approximate the transfer operator with a transition timescale $\tau$. We construct the reversibilized transition matrix $P_r$ and show the 15 slowest inferred relaxation times for $N=1000$ partitions and transition times $\tau$, Fig.\,\ref{fig_3_DW}(a-left). The longest relaxation time initially decays and reaches its asymptotic limit after $\tau \sim 5\,\text{s}$. As we increase $\tau$ to a few multiples of the hopping timescale, the eigenvalues stop changing and $t_\text{imp}$ simply grows linearly with $\tau$ (recall $t^\text{imp}(\tau)= \left|\Lambda^{-1}(\tau)\right| = -\tau/\log \lambda(\tau)$). Faster decaying eigenfunctions exhibit this behavior earlier. Importantly, when $t_\text{imp}$ is approximately constant the inferred timescales accurately predict the mean first passage time between potential wells, as expected from theory (see e.g. Ref. \onlinecite{VanKampen1992}). In fact, we find that the inferred $|\hat{\Lambda}_2|$ provides an excellent fit to the hopping rates across temperatures, Fig.\,\ref{fig_3_DW}(a-right). We note that in order to examine different temperatures the choice of $\tau^*$ should change accordingly to reflect the different resulting dynamics, Fig\,S5(b). Increasing the temperature reduces the hopping rates, and therefore, the amount of time it takes before the system relaxes to the stationary distribution. Nonetheless, the qualitative behavior of the longest inferred implied relaxation times is conserved across temperatures, Fig\,S5(b): a short $\tau$ transient quickly converges to the timescale of hopping between wells, before the system completely mixes and the implied timescales grow linearly. To choose $\tau^*$ consistently across temperatures we find the transition time that minimizes the longest reversibilized $t_\text{imp}(\tau)$. As expected the resulting $\tau^*$ is reduced as we increase the temperature, reflecting the faster dynamics.

To complement the above analysis of a stochastic dynamics,  we study the operator spectrum in the deterministic chaotic dynamics of the Lorenz system. Chaotic dynamics generally exhibit an intricate interplay of multiple timescales, which is apparent in the eigenvalue structure of $P_r$ for different transition times, Fig.\,\ref{fig_3_Lorenz}(a). Indeed, in the chaotic regime, the Lorenz dynamics meanders around a skeleton of unstable periodic orbits (UPOs) \cite{Farmer1987}, and that is reflected in the periodic behavior of the eigenvalues as a function of $\tau$. Nonetheless, for short timescales we find a regime in which the slowest implied timescale is approximately constant Fig.\,\ref{fig_3_Lorenz}(a-inset), and we choose $\tau^*=0.1\,\text{s}$ for subsequent analysis. Notably, despite the fact that we only have access to partial observations, the reconstructed Markovian dynamics are predictive over timescales of $\langle \tau_\text{pred} \rangle = 3.04\,(2.99,3.08)\,\text{lyapunov times}$, Fig.\,S7(a).

In both our example systems, the structure of the second eigenvector of the reversibilized transition matrix $\phi_2$ reveals a collective coordinate capturing transitions between coherent structures or almost-invariant sets, Fig.\,\ref{fig_3_DW}(b),\ref{fig_3_Lorenz}(b). To visualize the state space, we perform single value decomposition (SVD) on the reconstructed trajectory matrices $X_{K^*} = U\Sigma V^T$ and plot the projection onto the first two singular vectors $(v_1,v_2)$. Notably, in the double-well example the dominant SVD modes correspond to the position and velocity $(u_1,u_2)\sim (x, v)$ (see Fig.~S2(c)). A contour plot of the inferred $\phi_2$ for the double-well potential in the $(u_1,u_2)$ space shows that the sign of the eigenvector effectively splits the reconstructed state space into the two wells of the system, Fig.~\ref{fig_3_DW}(b). The same is true across temperatures, Fig.~S6, with the maximum of $\chi$, Eq.\,(\ref{Eq:coherence_min}), at $\phi_2^c\approx 0$ Fig.~S5(c). Notably, due to the underdamped nature of the dynamics the separation between potential wells includes the high velocity transition regions $|u_2|$, as expected from theory \cite{Risken1985}. Similarly, the sign of $\phi_2$ for the Lorenz system divides the state space into its almost-invariant sets, Figs.\,\ref{fig_3_Lorenz}(b),\,S7(b), which are partially split along the shortest period UPO \cite{Froyland2009, Brunton2017}, identified here through recurrences \cite{Lathrop1989} (Appendix \ref{sec:Methods}).

In summary, we introduce a general framework for the principled extraction of coarse-grained, slow dynamics directly from time series data, which applies to both deterministic and stochastic systems. To fully capture short-time dynamics we concatenate measurements in time and partition the expanded state while maximizing predictive information. We then leverage the geometric-invariance of the transfer operator representation to build an effective model of the dynamics from which we can extract coarse-grained metastable states.

\section{\label{sec:KS_entropy}Estimating the Kolmogorov-Sinai entropy}

Our approximation of the transfer operator dynamics yields an approximate model that holds for length scales and timescales beyond those set by the partition length scale $\epsilon \propto 1/N$ and the transition timescale $\tau$. However, certain fundamental properties of the underlying continuous dynamics are not immediately available from this representation. Among these, we here focus on the Kolmogorov-Sinai (KS) entropy, which is a fundamental measure of the ``compressibility'' of a dynamical machine \cite{Shannon1963}, capturing the inherent unpredictability of the dynamics. However important, the KS entropy is also extremely challenging to estimate from time series data \cite{Schurmann1996,Bollt2001}. The KS entropy is defined in the $\epsilon\rightarrow 0$, $\tau\rightarrow 0$ limit, and requires infinitesimal trajectory information to be properly estimated. This is reflected in our overestimation of the KS entropy in the Lorenz system, Fig.\,\ref{fig_2}(b), highlighting the challenge of bridging between a discrete representation and the underlying continuous dynamics. However, despite the fact that our transfer operator approximation is constructed for finite $\epsilon$ and $\tau$, we here show that we can establish a connection between our estimates of the short-time entropy rate in a partitioned space and the underlying KS entropy.

Our estimates of the short-time unpredictability of the dynamics through the entropy rate are related to the Shannon-Kolmogorov $(\epsilon,\tau)$-entropy per unit time \cite{Gaspard1993}, 

\begin{align}\label{eq:eps_tau_entropy_rate}
    h_{SK}(\epsilon,\tau) = \lim_{T\rightarrow\infty} \frac{1}{T} H_{SK}(\epsilon,\tau,T),
\end{align}
where $H_{SK}(\epsilon,\tau,T)$ is the Shannon-Kolmogorov $(\epsilon,\tau)$-entropy obtained from the probability that segments of length $T$ are within a distance smaller than $\epsilon$.  In comparison, our partition-based estimate of the short-time entropy rate, Eq.\,\ref{eq:h_p}, is equivalent to truncating the limit in Eq.~\ref{eq:eps_tau_entropy_rate} at $T=2\tau=2\delta t$ and setting $\epsilon\sim1/N$,

\begin{align*}
    h_{\delta t}(N,K) &= -\frac{1}{\delta t} \sum_{ij} \pi_i P_{ij}(\delta t)\log P_{ij}(\delta t) \\ 
    &\sim \frac{1}{2\delta t}H_{SK}(1/N,\delta t,2\delta t).
\end{align*}

Equivalently, we can define the partition-based Shannon entropy $H(S,\tau,T)$, where $S$ represents the symbol set composed of $N$ partitions, from which follows the definition of the Kolmogorov-Sinai entropy rate \cite{Kolmogorov1958,Gaspard1993},
\begin{align}\label{eq:KS_entropy}
    h_{KS} = \sup_S \lim_{\tau\rightarrow 0} \lim_{T\rightarrow\infty} \frac{1}{T}H(S,\tau,T),
\end{align}
which is the supremum of the entropy rate over all possible partitions of the state space. 

In practice, the KS entropy is extremely challenging to estimate from time series data, due to the several limits and the inherent arbitrariness of the choice of $\tau$ and partition $S$. The situation simplifies when the partition is generating \cite{Rokhlin1967,Bollt2013}, in which case there is a one-to-one mapping between state-space trajectories and the resulting symbolic dynamics, and therefore even a Markovian estimator of the entropy rate will correspond to the KS entropy. However, there are no general algorithms for finding generating partitions, with the exception of simple one dimensional discrete maps \cite{Schurmann1996} (e.g.~logistic map or tent map) or two dimensional maps for which heuristic approximation schemes have been successfully applied \cite{Grassberger1985,Kennel2003,Hirata2004} (e.g. Henon map or Ikeda map). A misplaced partition can result in the severe underestimation of the entropy \cite{Bollt2001}, whereas correlations that can stem from partial observations and/or an inappropriate partition result in the overestimation of the entropy \cite{Schurmann1996}. Given a finite partition, one can possibly circumvent the lack of a generating partition by taking the sequence length to infinity $T\rightarrow\infty$.  However, a na\"{i}ve estimator based on counting symbol sequences of increasing length $T$ is in practice unattainable: the number of possible symbolic sequences grows exponentially with the sequence length, making it impossible to reach moderate to large values of $T$ needed to reach the KS entropy limit. Indeed, this has motivated the development of algorithms for the estimation of entropy in undersampled symbolic sequences (see e.g. Refs. \onlinecite{Nemenman2001,Archer2013}).

Alternative geometric-based approaches, such as from Cohen and Procaccia \cite{Cohen1985}, overcome the issues of symbolic estimates of the entropy rate by working directly with the time series data and essentially estimating the probability of finding two trajectories of length $T$ within an $\epsilon$ distance of each other. Given a measurement time series $\{y(t)\}$ where $t\in \{\tau,2\tau,\ldots,T\}$, $\epsilon$-tubes are built around sequences of length $K\tau$, $y_{t:t+K}$, and the entropy is estimated as,

\begin{equation}
    H^C_{K\tau}(\epsilon) = -\left\langle \text{log} C_{K\tau}\left(\frac{\epsilon}{2}\right) \right\rangle,
\end{equation}

\noindent where $C_{K\tau}\left(\frac{\epsilon}{2}\right)$ is the correlation function of $K$-dimensional $\epsilon$-tubes built from time series measurements with a sampling time of $\tau$, which essentially measures the probability of finding two $y_{t:t+K}$ vectors within a distance $\epsilon$ of each other. This notion allows for the estimation of a number of ergodic properties of the dynamics, including intrinsic dimensions and entropies. For example, the correlation entropy rate can then be obtained by estimating,

\begin{equation}\label{eq:h_c}
    h^C_\tau(\epsilon) = \lim_{K\rightarrow\infty}\left[ H^C_{(K+1)\tau}(\epsilon) - H^C_{K\tau}(\epsilon)\right],
\end{equation}

\noindent which converges to the Kolmogorov-Sinai entropy in the limit $\epsilon\rightarrow 0$ and $\tau \rightarrow 0$. Alternatively, for some classes of dynamical systems one can use the sum of positive Lyapunov exponents to obtain an upper bound to the KS entropy, according to Pesin's theorem \cite{Pesin1977}. However powerful, geometric approaches can be challenging to apply directly to data, since they are sensitive to the underlying geometry and the precise dimension of the reconstructed state space. 

\begin{figure}[ht!]
\begin{center}
\includegraphics[scale=1]{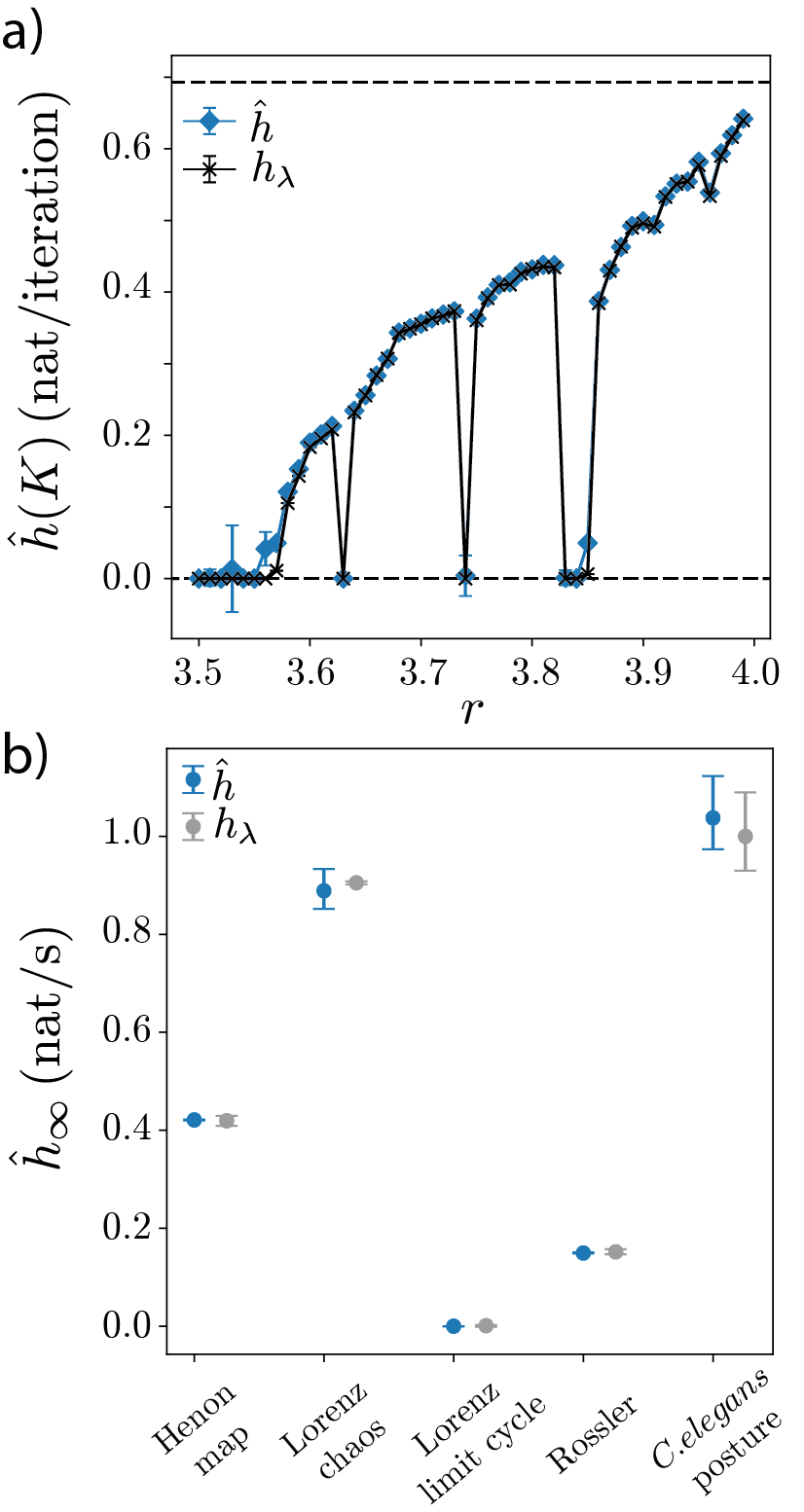}
\caption{{\bf Approaching the Kolmogorov-Sinai (KS) entropy in a partitioned state space.}
(a) KS entropy estimates (blue) in the logistic map with varying control parameters $r$ and the corresponding Lyapunov exponent $h_\lambda$ (black crosses), spanning qualitatively different dynamical regimes (details in Appendix \ref{sec:Methods_KS_entropy}). Black horizontal dashed lines denote the extremes for the KS entropy in the logistic map: $h_\lambda=0$ (non-chaotic dynamics) and $h_\lambda = \log 2$ (maximally chaotic dynamics).
(b) Estimates of the KS entropy across multiple systems. Our estimates $\hat{h}$ (blue) are in excellent agreement with the direct estimation of the sum of positive Lyapunov exponents\cite{Pesin1977} $h_\lambda$ (gray). 
}
\label{fig4}
\end{center}
\end{figure}

Here we present a novel estimator of the KS entropy, which combines ideas from previous approaches with the transfer operator formalism, to obtain accurate estimates of the KS entropy that are robust to the precise geometry of the reconstructed state space and do not require a generating partition. As Cohen and Procaccia \cite{Cohen1985}, we build trajectory segments of increasing length $T = K\delta t$, but then partition the resulting space and estimate the Markov approximation of the transfer operator for $\tau = K\delta t$,

\begin{align}\label{eq:h_KS(K,N)}
    \hat{h}(N,K) =  -\frac{1}{K\delta t} \sum_{ij} \pi_i P_{ij}(K\delta t)\log P_{ij}(K\delta t).
\end{align}
We then estimate the KS entropy as a supremum over $N$, $\hat{h}(K) =\max_N \hat{h}(N,K)$, estimate $\hat{h}(K)$ for increasing number of delays and probe the $K\rightarrow\infty$ limit (see Appendix \ref{sec:Methods_KS_entropy},

\begin{align}\label{eq:h_KS(K)}
    h_{KS} = \lim_{K\rightarrow\infty}\sup_N \hat{h}(N,K).
\end{align}

In Fig.\,\ref{fig4} we illustrate our estimation procedure with applications in discrete maps (logistic and Henon), and in the continuous-time dynamics of the Lorenz system and the R\"{o}ssler system, as well as a real world data example from \emph{C. elegans} locomotion, where a complementary estimate of the entropy rate derived from Lyapunov exponents already exists \cite{Ahamed2021}.  We assess the accuracy of our KS entropy estimates by comparing against the sum of positive Lyapunov exponents \cite{Pesin1977}. For the discrete maps, we find that the direct estimates of $\hat{h}(K)$ provide an accurate approximation of the KS entropy, provided that we choose $K$ before finite-size effects become evident, as shown in Fig.\,S8(a). In contrast, for continuous-time dynamics we find that it is often challenging to reach the KS entropy limit directly and we instead estimate the $K\rightarrow\infty$ limit by extrapolation \cite{Strong1998}, Fig.\,S8(b) (Appendix \ref{sec:Methods_KS_entropy}).  In both discrete and continuous systems, our entropy estimates from Eq.~\ref{eq:h_KS(K)} are in excellent agreement with the sum of positive Lyapunov exponents.

\section*{Discussion}

When seeking order in complex systems, many simplifications are possible, and this choice has consequences for both the nature and complexity of the resulting model. Often such simplifications are chosen {\em a priori},  for example a particular discretization, and the complexity of the model is then taken as a fact of the underlying dynamics. But this is not generally true; the definition of the states and the resulting dynamics between them are inextricably linked. Here we address this link directly and choose the simplification informed by the resulting model dynamics; we define maximally predictive states in an attempt to achieve an effective Markov model of the large-scale dynamics. Applicable to both deterministic and stochastic systems, we combine state-space reconstruction with ensemble dynamics to infer probabilistic state space transitions and coarse-grained descriptions directly from data.

By characterizing nonlinear dynamics through transitions between state-space partitions, we trade trajectory-based techniques for the analysis of linear operators. This approach is complementary to previous methods of state-space reconstruction which focus on geometric and topological quantities \cite{Packard1980,Farmer1982,Pei1996,Lathrop1989,Ahamed2021}. In particular, the ergodic analyses of trajectories rely on precise estimates of dimension and of the local Jacobian, which is challenging in systems with unknown equations. 

In our approach, we recognize and address the mutual dependence of state-space reconstruction and Markovian evolution: reconstructing the state space is required for an effective Markovian description and the framework of ensemble dynamics provides the principled, information-theoretic measure of memory used to optimize the reconstructed state. By reconstructing the state space, hidden dynamics are included into the state, such that the slow modes uncovered by the operator reconstruction correspond to the slowest possible ergodic dynamics in the system. 

The approximation of transfer operators for the extraction of long-lived states has been successfully applied in a number of systems: from identifying metastable protein conformations in molecular dynamics simulations to determining coherent structures in oceanic flows (see e.g. Refs. \onlinecite{Schutte2001,Froyland2007,Dellnitz2009,Bollt2011,Bollt2013,Froyland2014,Chodera2014,bowman_introduction_2014,Klus2018,Fackeldey2019}). Notably, most of these approaches benefit from a detailed physical understanding of the systems under study, which can be leveraged to approximate the transfer operator with high precision. However, finding optimal coordinates and metrics that can capture kinetic differences between metastable states is an active field of research \cite{Chodera2014}. Interestingly, recent advances have been achieved by using temporal information to find better coordinates for Markov state modeling \cite{Perez-Hernandez2013,Schwantes2013,Wang2016}, which results from an effective delay-embedding of the underlying state-space dynamics. In our approach, representation and modeling are unified under a single framework: we search for maximally predictive state-space reconstructions that yield a representation in which the transfer operator dynamics optimally captures the slow dynamics of the system and, thus, can appropriately identify timescale separation and metastable states from incomplete time series measurements. 

To identify almost-invariant sets, we use the reversibilized operator $P_r$, Eq.\,(\ref{eq:rev_P}), which is generally different from $P$. For an overdamped system in thermodynamic equilibrium, this symmetrization simply enforces detailed balance and $|\Lambda_2^{-1}|$ is directly related to the hopping rates between metastable states. For irreversible dynamics, however, the dynamics of $P_r$ will relax more slowly than those of $P$, providing an upper bound to the true relaxation times \cite{Fill1991,Ichiki2013}.  It will also be interesting to explore approaches based directly on $P$ \cite{Fackeldey2019}, and to extract additional physical quantities such as the rate of entropy production.

We restrict our analysis to autonomous ergodic systems, but slow non-ergodic variables that drive the dynamics on timescales comparable to the observation time can render the data non-stationary. In this case, the transfer operator has an explicit time dependence $\mathcal{P}_\tau(t)$ that we cannot capture within our current approach. Nonetheless, varying $\tau$ might still allow for the identification of coherent sets or collective coordinates that dominate the dynamics on the timescale $\tau$\cite{Froyland2010,Froyland2010b,Wang2015,Koltai2016,Koltai2018,Fackeldey2019}.

When the state space is known or its reconstruction decoupled from the ensemble dynamics, mesh-free discretizations have been used to characterize ensemble evolution,  including diffusion maps \cite{Nadler2006,Brennan2019} and other kernel-based approaches, such as Reproducing Kernel Hilbert Spaces \cite{Klus2020} or Extended Dynamic Mode Decomposition \cite{Klus2016}. Though powerful, such methods require subtle choices in kernels, neighborhood length scales and transition times, for which we lack guiding principles. Nonetheless, incorporating such approaches into our framework would likely be fruitful. In particular, when the measurement time series is high dimensional our partitioning scheme will likely struggle: for large $K$ the trajectory matrices are high dimensional, making it challenging to perform k-means clustering. In this case, a prior dimensionality reduction step or a more appropriate choice of metric might be required.

The principled integration of fluctuating, microscopic dynamics resulting in coarse-grained but effective theories is a remarkable success of statistical physics. Conceptually similar, our work here is designed toward systems sampled from data whose fundamental dynamics are unknown.  We leverage, rather  than ignore, small-scale variability to subsume nonlinear dynamics into linear, ensemble evolution, enabling the principled identification of coarse-grained, long-timescale processes, which we expect to be informative in a wide variety of systems. 

\section*{Supplementary Material}

See the supplementary material for Figs. S1-S9.

\begin{acknowledgments}
    We thank Massimo Vergassola and Federica Ferretti for comments and lively discussions.  This work was supported by  a program grant from the Netherlands Organization for Scientific Research (Sticthing voor Fundamenteel Onderzoek der Materie): FOM V1310M (AC, GJS),   by the LabEx ENS-ICFP: ANR-10-LABX-0010/ANR-10-IDEX-0001-02 PSL (AC) and we also acknowledge support from the OIST Graduate University (TA, GJS), the Herchel Smith Fund (DJ), and the Vrije Universiteit Amsterdam (AC, GJS). GJS and AC acknowledge useful (in-person!) discussions at the Aspen Center for Physics, which is supported by National Science Foundation Grant PHY-1607611.
\end{acknowledgments}

\section*{Author contributions}
All authors contributed equally to this work.

\section*{Data Availability Statement}

The data that support the findings of this study are openly available in Zenodo at \url{https://doi.org/10.5281/zenodo.7130012}. In addition, code for reproducing our results is publicly available:
\url{https://github.com/AntonioCCosta/maximally_predictive_states}.

\appendix

\section{\label{sec:Methods} Methods}

\medskip

\noindent{\bf State-space reconstruction:} Given a measurement time series, $\vec{y}(t)$, with $t\in\{\delta t,\ldots,T \delta t\}$ and $\vec{y}\in \mathbb{R}^d$, we build a trajectory matrix by stacking $K$ time-shifted copies of $\vec{y}$, yielding a $(T-K)\times Kd$ matrix $X_K$. For each $K$, we partition the candidate state space and estimate the entropy rate of the associated Markov chain (see below). We choose $K^*$ such that $\left. \partial_K h_{\delta t}(K,N^*)\right|_{K^*} \sim 0$.

\medskip

\noindent{\bf State-space partitioning:} We partition the state space into $N$ Voronoi cells, $s_i, i\in\{1,\ldots,N\}$, through k-means clustering with a k-means++ initialization using scikit-learn \cite{scikit-learn}.

\medskip

\noindent{\bf Approximation of the Perron-Frobenius operator:} We build a finite-dimensional approximation of the Perron-Frobenius operator using Ulam-Galerkin discretization. A Galerkin projection takes the infinite-dimensional operator onto an $N\times N$ operator of finite rank by truncating an infinite-dimensional set of basis functions at a finite $N$. Ulam's method uses characteristic functions as the basis for this projection,

\begin{equation}\label{eq:UlamBasis}
    \zeta_i(x) = \begin{cases}
    1, & \text{for $x \in s_i$}\\ 
    0, & \text{otherwise}
    \end{cases}.
\end{equation}

\noindent Our characteristic functions are implicitly defined through the k-means discretization of the space. We thus partition the space into $N$ connected sets with an nonempty and disjoint interior that covers $M$: $M = \cup_{i=1}^N s_i$, and approximate the operator as a Markov chain by counting transitions from $s_i$ to $s_j$ in a finite time $\tau$. Given T observations, a set of $N$ partitions, and a transition time $\tau$, we compute

\begin{equation}
    C_{ij}(\tau) = \sum_{t=0}^{T-\tau}\zeta_i(x(t))\zeta_j(x(t+\tau)). \nonumber
\end{equation}
The maximum likelihood estimator of the transition matrix is obtained by simply row normalizing the count matrix,

\begin{equation}\label{eq:Pij}
    P_{ij}(\tau) = \frac{C_{ij}(\tau)}{\sum_j C_{ij}(\tau)}. \nonumber
\end{equation}

\medskip

\noindent{\bf Invariant density estimation:} Given a transition matrix $P$, the invariant density is obtained through the left eigenvector of the non-degenerate eigenvalue 1 of $P$, $\pi P = \pi$.

\medskip

\noindent{\bf Short-time entropy rate estimation:} Given a transition matrix $P(\tau=\delta t)$ and its corresponding invariant density $\pi$ we compute the short-time entropy rate through Eq.~(\ref{eq:h_p}).

\medskip

\noindent{\bf Markov model simulations:} At each iteration, we sample from the conditional distribution $P(s_j(t+\delta_t)|s_i(t))$, given by the $i$-th row of the inferred $P_{ij}(\delta t)$ matrix, to generate a symbolic sequence. At each frame $t$, we then randomly sample a state-space point within the partition corresponding to the simulated symbol and unfold it to obtain $x_{t-K^*/2:t+K^*/2}$, from which we get $x_\text{sim}(t)$.

\medskip

\noindent{\bf Metastable state identification:} Metastable states correspond to regions of the state space that the system visits often, separated by regions where transitions occur. We therefore search for collections of states that move coherently with the flow: their relative future states belong to the same macroscopic set. We leverage a heuristic introduced in Ref.~\onlinecite{Dellnitz1999}, which makes use of the first non-trivial eigenfunction of the reversibilized transfer operator to identify almost-invariant sets \cite{Froyland2009}: the coherence properties are invariant to this transformation \cite{Froyland2005} and the analysis is simplified due to the optimality properties of reversible Markov chains. In discrete time and space, $P_r$ is defined as

\begin{equation}
    P_r(\tau) = \frac{P(\tau)+P(-\tau)}{2},
\label{eq:rev_P}
\end{equation}

\noindent where, 
\begin{equation}
    P_{ij}(-\tau) = \frac{\pi_j P_{ji}(\tau)}{\pi_i}. \nonumber
\end{equation}

\noindent is the stochastic matrix governing the time reversal of the Markov chain. The first nontrivial right eigenvector of $P_r$, $\phi_2$, allows us to define macrostates as

\begin{equation}
    S^+(\phi_2^c) \coloneqq \bigcup_{i:\phi_2\geq \phi_2^c} s_i\,\\,\,S^-(\phi_2^c) \coloneqq \bigcup_{i:\phi_2<\phi_2^c} s_i,
\end{equation}

\noindent and $\phi_2^c$ is chosen so as to maximize $\chi$, Eq.\,(\ref{Eq:coherence_min}), yielding almost-invariant or metastable states. In practice, we compute Eq.\,(\ref{Eq:coherence_min}) for the complete discrete set of $\phi_2^c$ and find the global maximum: this is an inexpensive calculation since $\chi$ can be obtained by matrix multiplications and $\phi_2^c$ can only take $N$ different values, where $N$ is the number of partitions. We note that for an overdamped thermodynamic system in equilibrium, the eigenvectors $\phi_k$ are discrete approximations of the eigenfunctions of the Koopman operator \cite{Klus2016}, which are nearly constant within a metastable set thus simplifying the clustering into almost-invariant sets.

\medskip

\noindent {\bf Choice of transition time $\tau^*$:} We choose $\tau^*$ as the shortest transition timescale after which the inferred implied relaxation times reach a plateau. For $\tau$ too short, the approximation of the operator will yield a transition matrix that is nearly identity (due to the finite size of the partitions and too short transition time), which results in degenerate eigenvalues close to $\lambda\sim 1$: an artifact of the discretization and not reflective of the underlying dynamics. For $\tau$ too large, the transition probabilities become indistinguishable from noisy estimates of invariant density, which results in a single surviving eigenvalue $\lambda_1=1$ while the remaining eigenvalues converge to a noise floor resulting from a finite sampling of the invariant density. Between such regimes, we find a region with the largest time scale separation (as illustrated in Fig.\,\ref{fig_1}(b-middle)) which also corresponds to the regime for which the longest relaxation times $t_\text{imp}$, Eq.\,(\ref{eq:t_imp}), are robust to the choice of $\tau$. We compute $t_\text{imp}$ using the eigenvalues of the reversibilized transition matrix, $P_r$, which only gives an upper bound to the relaxation dynamics. 

\medskip

\noindent {\bf Eigenspectrum estimation:} Since we are focused on long-lived dynamics, we estimate the $n_\text{modes}$ largest magnitude real eigenvalues using the ARPACK \cite{ARPACK} algorithm.

\medskip

\noindent{\bf Periodic Orbit identification:} We identify the shortest period unstable periodic orbit of the Lorenz system, Eq.\,(\ref{eq:Lorenz}), by studying the distribution of recurrence times. We set a short distance $\epsilon$ and look for the times $p$ at which $||X_{t+p}-X_{t}||<\epsilon$, where $||\cdot||$ represents the Euclidean distance. In practice, we compute $1/||X_{t+p}-X_{t}||^2$ and find peaks with height larger than $1/\epsilon^2$ using the {\tt find\_peaks} function from the scipy.signal package \cite{Scipy}. For short enough $\epsilon$ ($\epsilon = 5\times 10^{-6}$ in our case), the distribution of recurrence times has its first peak at the period of the shortest unstable periodic orbit \cite{Barrio2015}. A trajectory corresponding to a shadow of this unstable periodic orbit \cite{Hammel1987,Nuse1988} is shown in Fig.\,\ref{fig_3_Lorenz}(b) and Fig.\,S7(b).

\section{\label{sec:Simulations} Simulations}

\noindent{\bf Double-well:} We use an Euler-Maruyama integration scheme to simulate Eq.\,(\ref{eq:DW_phspace}) and generate a $T = 10^6\text{s}$ long trajectory of a particle in a double-well potential with $m=\gamma=1$ and $\beta^{-1} = \{0.5,0.75,1.00,1.25,1.5,1.75,2.00,2.25,2.5\}\,\text{J}$. We first sampled at $0.01 \,\text{s}$, and then downsampled to $\delta t = 0.05\,\text{s}$. In addition, the first $1000\,\text{s}$ were discarded to avoid transients. In Fig.~S5(a), we project the Boltzmann distribution into the SVD space by learning a linear mapping $\theta$, between $[u_1,u_2]$ and $[x,v]$: $[u_1,u_2] = \theta\cdot [x,v]$. This allows us to project the centroids of each Voronoi cell from the $[u_1,u_2]$ space to the $[x,v]$ space, and estimate the corresponding Boltzmann weight $p([u_1,u_2]\cdot\theta^{-1}) = p([x,v]) = \exp\{-\beta\left[(x^2-1)^2+v^2/2\right]\}/Z$, where $Z$ represents the normalization.

\medskip 

\noindent{\bf Lorenz system:} We use scipy's {\tt odeint} package \cite{Scipy} to generate a $T =5\times 10^5\,\text{s}$ long trajectory of the Lorenz system, Eq.\,\ref{eq:Lorenz}, in the standard chaotic regime $(\sigma,\rho,\beta) = (10,28,8/3)$, sampled at $\delta t = 0.01 \,\text{s}$. We discard the first $10^3\,\text{s}$ to avoid transients.

\medskip

\section{\label{sec:Methods_KS_entropy} Estimating the Kolmogorov-Sinai entropy rate in a partitioned state space}

In order to estimate the KS entropy of the source, we build trajectory matrices for increasing delays $K$, $X_K$, and estimate the entropy rate for increasing $K$ on a transition time $K\delta t$, Eq.\,\ref{eq:h_KS(K,N)}. In order to capture as much information as possible in the partitioned state space, we take the supremum over $N$, $\hat{h}(K) =\max_N \hat{h}(N,K)$, and probe the $K\rightarrow\infty$ limit Eq.\,\ref{eq:h_KS(K)}. We note that Eq.\,\ref{eq:h_KS(K)} has an implicit $\delta t$-dependency, which we lift by probing the consistency of the estimation for different values of $\delta t$, as exemplified in the R\"{o}ssler system, Fig.\,S9. We aim for $\delta t\rightarrow 0$, while paying attention to the fact that when $\delta t$ is too short, we need longer $K$ to reach the $T=K\delta t\rightarrow \infty$ limit, which might be impractical. We use different methods to probe the $K\rightarrow\infty$ limit in discrete maps and in continuous-time dynamical systems. For discrete maps, we find that it is possible to achieve an accurate estimator of the KS entropy with a short amount of delays, and we identify the $K\rightarrow\infty$ as the largest $K$ before which finite-size effects start becoming evident. Typically, as $K$ grows, the change in $\hat{h}(K)$ with $K$ slows down (typically as $~1/K$), meaning that 

\begin{align}\label{eq:delta_h}
    \Delta \hat{h}(K) = \hat{h}(K)-\hat{h}(K+1),
\end{align}
should be a monotonically decreasing function. However, we find that finite-size effects result in a underestimation of the entropy (beyond the expected $\sim 1/K$), which results in a increase in $\Delta \hat{h}(K)$. Therefore, we identify the largest $K$ before finite-size effects become significant by taking the minimum of $\Delta \hat{h}(K)$, Fig.\,S8(a). In comparison, we find that given our choice of $\delta t$ for the continuous-time dynamical systems studied here, the value of $K$ required for an accurate estimate of the KS entropy is unattainable. In order to reach the $K\rightarrow\infty$ limit, $h_\infty$, we leverage the $1/K$ behavior of $\hat{h}(K)$ to extrapolate to $K\rightarrow\infty$ by weighted least squares regression of $\hat{h}(K) = m/K+h_\infty$ as in Ref.~\onlinecite{Strong1998}, Fig.\,S8(c). We note that this observation is fundamentally tied to the choice of sampling time $\delta t$ as discussed in Fig.\,S9.

\section{\bf Additional simulations for Sec.\,\ref{sec:KS_entropy}}

\noindent{\bf R\"{o}ssler system}: We use Scipy's {\tt odeint} package \cite{Scipy} to generate a $T =5\times 10^3\,\text{s}$ long trajectory of the R\"{o}ssler system\cite{Rossler1976,Peitgen2004},
\begin{align}\label{eq:Rossler}
    \begin{cases}
    \dot{x} = & -y-z  \\
    \dot{y} = & x+ay\\
    \dot{z} = & b+z(x-c)
    \end{cases},
\end{align}
in a chaotic regime with $(a,b,c) = (0.52, 2, 4)$, sampled at $\delta t = 0.1 \,\text{s}$. We discard the first $100\,\text{s}$ to avoid transients.

\medskip

\noindent{\bf Lorenz system in limit cycle regime}: We use scipy's {\tt odeint} package \cite{Scipy} to generate a $T = 1000\,\text{s}$ long trajectory of the Lorenz system, Eq.\,\ref{eq:Lorenz}, in a limit cycle regime with $(\sigma,\rho,\beta) = (10,300,8/3)$, sampled at $\delta t = 0.01 \,\text{s}$. We discard the first $100\,\text{s}$ to avoid transients.

\medskip

\noindent{\bf Henon map}: We iterate the Henon map \cite{Henon1976},

\begin{align}\label{eq:Henon}
    \begin{cases}
    x_{t+1} = & a+by_t-x_t^2 \\
    y_{t+1} = & x_t
    \end{cases},
\end{align}
in a chaotic regime $(a,b) = (1.4,0.3)$, to generate a trajectory with $T=10^6$ time steps, discarding the first $1000$ iterations to avoid transients. 

\medskip

\noindent{\bf Logistic map}: We iterate the logistic equation,

\begin{align}\label{eq:Logistc}
    x_{t+1} =  rx_t(1-x_t)
\end{align}
to generate a trajectories with $T=10^6$ time steps, discarding the first $100$ iterations to avoid transients. We change the parameter $r$ to span multiple dynamical regimes in the period-doubling route to chaos.

\section*{References}

\bibliography{Bibliography}

\clearpage

\onecolumngrid

\setcounter{figure}{0}
\setcounter{page}{1}

\makeatletter
\renewcommand{\theequation}{S\arabic{equation}}
\renewcommand{\thefigure}{S\arabic{figure}}

\section*{Supplementary Material}

\begin{figure*}[ht!]
\begin{center}
\includegraphics[scale=1]{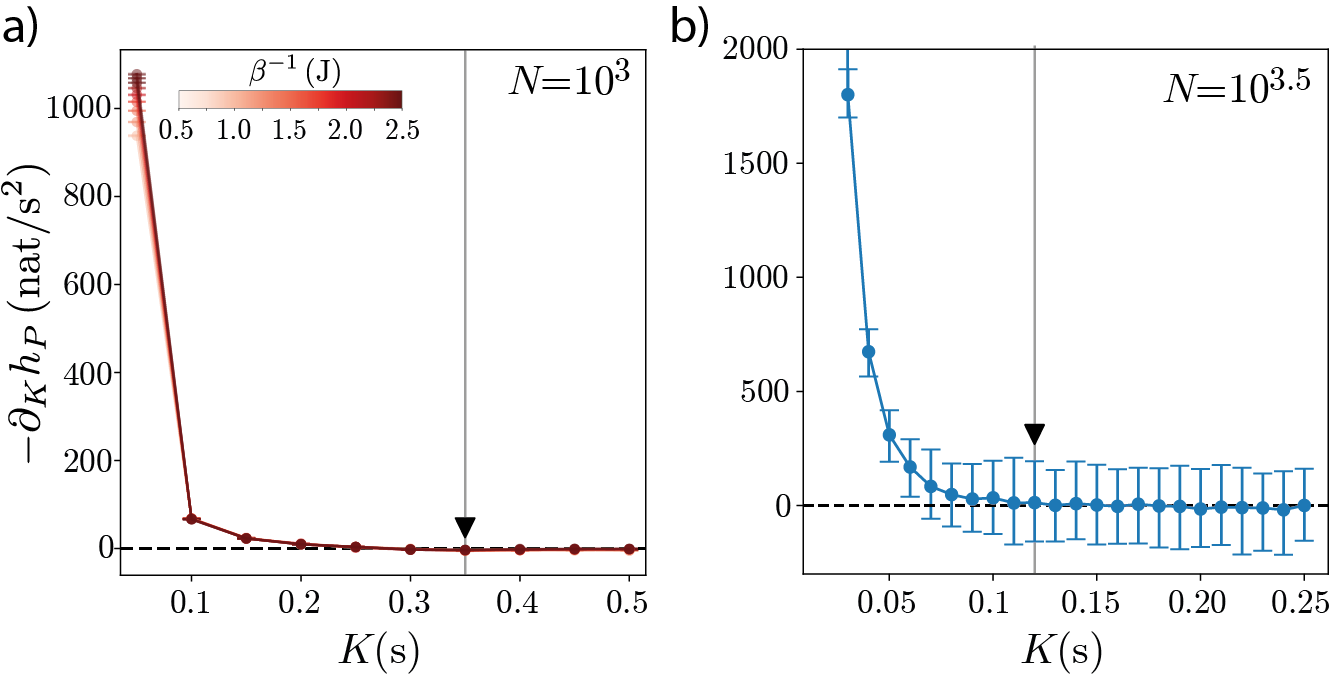}
\caption{{ \bf Change in entropy rates as a function of delays $K$ for the particle in the double-well and the Lorenz system.}
(a) Double-well dynamics for $N=10^{3}$ partitions. Across all temperatures sampled here, the entropy rate reaches a plateau after $K\gtrsim 6\,\text{frames} = 0.3\,\text{s}$ and we choose $K^* = 7\,\text{frames} = 0.35\,\text{s}$. Notably, the change in entropy rate is still significantly nonzero at $K=2\,\text{frames}$, reflecting the non-Markovian effects of the Euler integration scheme as discussed in the main text. Error bars are 95\% confidence intervals bootstrapped over $50000\,\text{s}$ trajectory segments.
(b) Lorenz system for $N=10^{3.5}$ partitions. The entropy rates reaches a plateau after $K\gtrsim 0.1\,\text{s}$ and we choose $K^* = 0.12\,\text{s}$. Error bars represent 95\% confidence intervals bootstrapped over $2000\,\text{s}$ trajectory segments.
}
\label{fig_S_h_diff}
\end{center}
\end{figure*}

\begin{figure*}
\begin{center}
\includegraphics{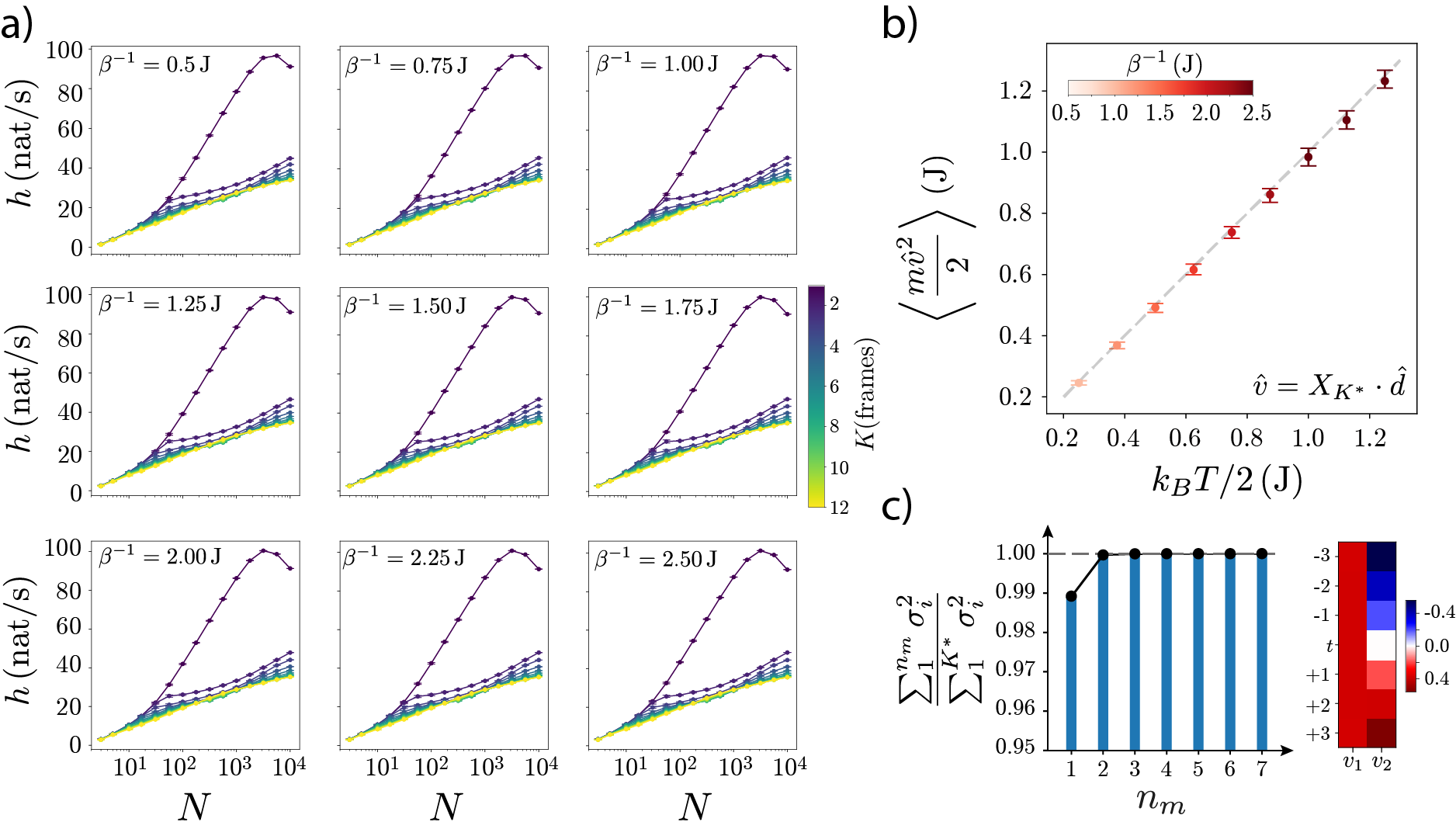}
\caption{{\bf Details of the state space reconstruction of an underdamped particle in a double-well potential.}
(a) Entropy rate as a function of delay $K$ and number of partitions $N$ across temperatures. The qualitative behavior of the entropy rate is conserved across temperatures.
(b) We recover the equipartition theorem in the reconstructed state space, an indication that complete information is contained in the reconstructed state space. We use a 6th-order finite difference approximation to estimate the velocity $\hat{v} = X_{K^*}\cdot \hat{d}$ from the reconstructed state $X_{K^*}$ with $K^* = 7\,\text{frames}$, where $\hat{d}$ represent the 6th-order finite difference stencil. We then compute the average kinetic energy and show that it closely matches the expected value of $(k_B T)/2$ from equipartition. We use $K^* = 7\,\text{frames}$ across temperatures. Error bars are 95\% bootstrapped confidence intervals of the mean velocity bootstrapped over $500\,\text{s}$ time points.
(c) Singular value decomposition (SVD) of the reconstructed state space $X_K^*=U\Sigma V^T$ with $K^* = 7\,\text{frames}$. The first two SVD modes $(u_1,u_2)$ capture nearly all of the variance as measured through the singular values $\sigma_i$ (c-left). The corresponding singular vectors capture the position and the velocity (c-right): $v_1$ is a weighted sum of positions, while $v_2$ is a derivative filter, yielding an estimate of the local velocity.}
\label{fig_S_DW_SVD}
\end{center}
\end{figure*}

\begin{figure*}
\begin{center}
\includegraphics{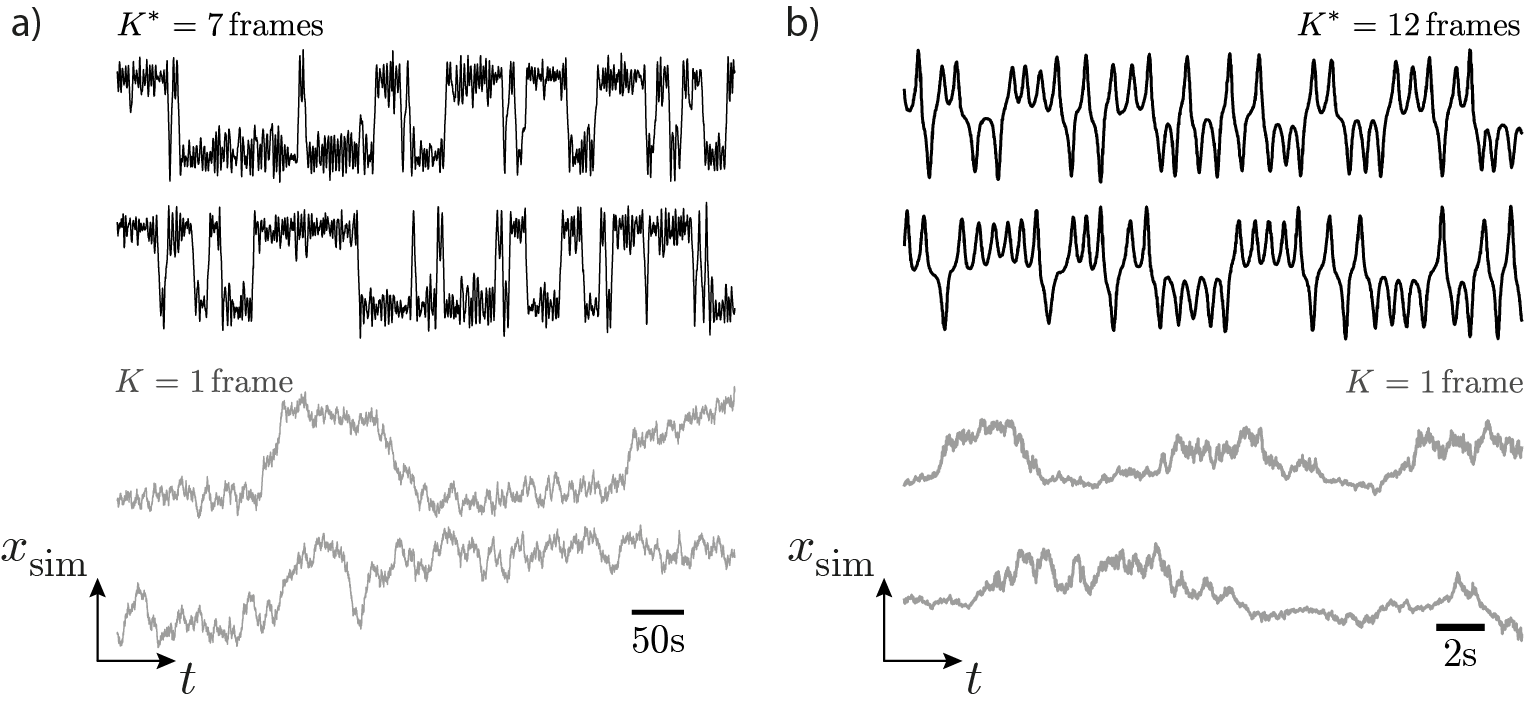}
\caption{{\bf Transfer operator simulations of the measured variables in the double-well and the Lorenz system.}
We leverage the inferred Markov chain $P_{ij}(\delta t)$ to simulate analog measurement time series. At each iteration we sample the next symbol according to the conditional probability distribution given by the $i$-th row of the inferred $P_{ij}$ matrix, and then randomly sample a point in the corresponding $X_K$ space from which we can recover a simulated time series $x_\text{sim}$ (Appendix \ref{sec:Methods}).
Example simulations of the double-well potential (a) and the Lorenz system (b) obtained with $K^*$ (top, black) and $K=1\,\text{frame}$ (bottom, gray).
}
\label{fig_S_Markov_sims}
\end{center}
\end{figure*}

\begin{figure*}
\begin{center}
\includegraphics[scale=1]{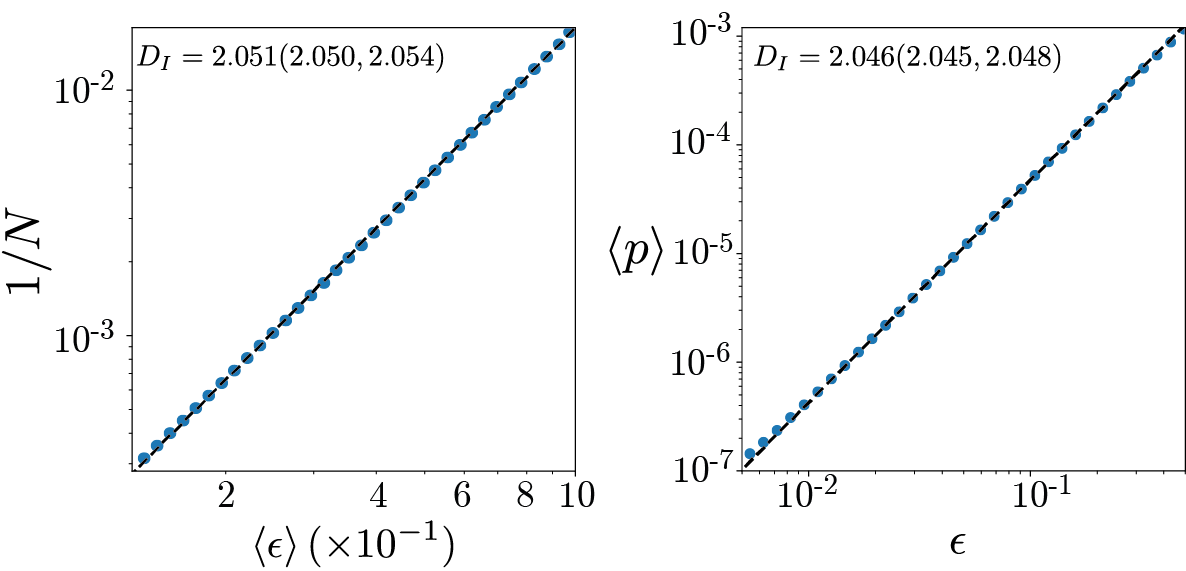}
\caption{{\bf Estimation of information dimension through transfer operator dynamics in the Lorenz system.}
In a typical chaotic attractor, the average probability $\langle p\rangle$ of finding a neighbor within a length scale $\epsilon$ scales as $\langle p \rangle = \epsilon^{D_I}$, where $D_I$ is the information dimension \cite{Farmer1983} in the limit $\epsilon\rightarrow 0$.
(left) We estimate the information dimension in a partitioned reconstructed state space, with $K^*=12\,\text{frames}$ Fig.\,\ref{fig_2}(b). The probability of belonging to a partition scales with the inverse of the number of partitions, and thus $p \propto 1/N \propto \epsilon^{D_I}$. Therefore, we can directly estimate the information dimension by measuring how the within cluster distances scale with increasing $N$. We estimate $D_I$ by least-squares regression of $\log N = -D_I \log\langle\epsilon\rangle + c$, where $\langle\cdot\rangle$ represents the median and $c\in \mathbb{R}$ is a constant. Error estimates are 95\% confidence intervals bootstrapped across $5000\,\text{s}$ segments; error bars are smaller than the marker size.
(right) Geometric estimation of the correlation dimension using the Cohen-Procaccia method \cite{Cohen1985} in the Lorenz system. Given the measurements of the $x$ variable of the Lorenz system, we build 50000 random sequences $x_{t:t+K^*}$ of $K^*=12\,\text{frames}$ and count the number of pairs of sequences within a distance $\epsilon$ from each other using a Chebyshev distance. $D_I$ is then estimated by least-squares regression of $\log\langle p\rangle\propto D_I \log\epsilon$. Error estimates are 95\% confidence intervals bootstrapped across 50000 randomly sampled sequences.
}
\label{fig_SLorenz_corr_dim}
\end{center}
\end{figure*}

\begin{figure*}
\begin{center}
\includegraphics{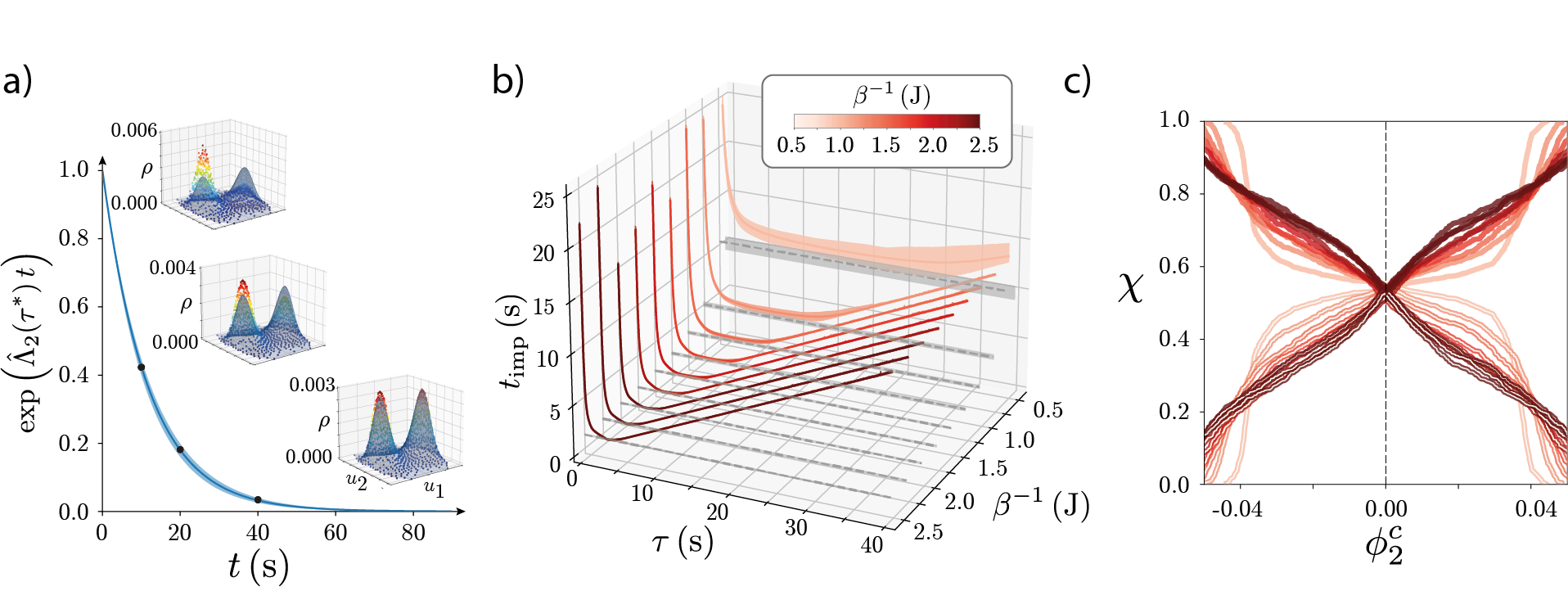}
\caption{{\bf Transfer operator dynamics in the double-well potential and metastable state identification across temperatures.}
(a)  We simulate the ensemble dynamics using the inferred transfer operator $\rho(t+\tau^*) = P(\tau^*)\rho(t)$, starting from an initial ensemble sharply concentrated on a single partition on the bottom of the left well, for $\beta^{-1}=0.5\,\text{J}$, $N=1000$ partitions and $\tau^* = 10\,\text{s}$. The decay of the ensemble dynamics projected onto $\phi_2$ captures the slow relaxation towards the equilibrium density. In the inset, we show the density at three different snapshots of the dynamics (black dots), projected onto the two largest singular vectors of the reconstructed state space $(u_1, u_2)$,  which match the position and velocity degrees of freedom, Fig.\,\ref{fig_S_DW_SVD}(c).
(b) Inferred longest implied time scales $t_\text{imp}$ as a function of transition time $\tau$ and effective temperature $\beta^{-1}=k_B T$. The qualitative behavior of $t_\text{imp}$ with $\tau$ is conserved across temperatures: early transient decay to the time scale of the hopping rate $t_\text{MFP}/2$ (gray dashed line) and long time linear increase with $\tau$ indicating independence of the states. We leverage the conserved qualitative behavior of $t_\text{imp}$ across temperatures to choose $\tau^*$ as the time scale that minimizes $t_\text{imp}$. This results in a lower $\tau^*$ for higher temperatures, which reflects the faster nature of the dynamics.
(c) Coherence measure $\chi$, Eq.\,(\ref{Eq:coherence_min}), as a function of the threshold in $\phi_2$ used to define the metastable states, $\phi_2^c$, for different temperatures. For each temperature, we show both $\chi_{\mu,\tau}(S^+)$ and  $\chi_{\mu,\tau}(S^-)$ as well as the minimum between them, $\chi$, in a gray dashed line. Changing $\tau^*$ to reflect the nature of the dynamics yields a relatively constant maximum $\chi$ across temperatures. Notably, at $\tau^*$ the coherence measure indicates that nearly half the measure has escaped the metastable states.
}
\label{fig_S_DW_coarse_graining}
\end{center}
\end{figure*}

\begin{figure*}
\begin{center}
\includegraphics{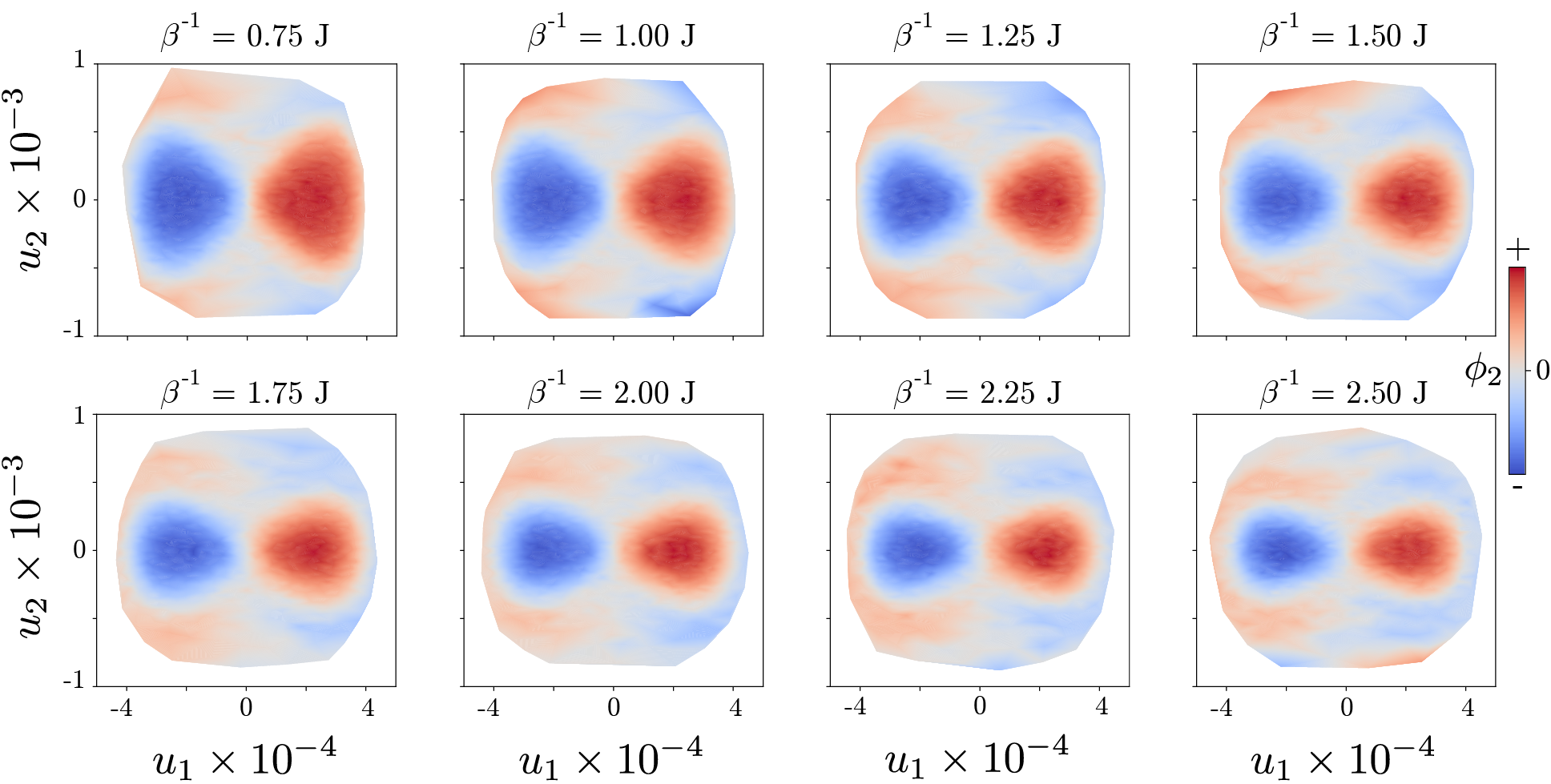}
\caption{{\bf Inferred slow eigenfunction of the double-well potential across temperatures.}
Contour plot of $\phi_2$ projected onto the first two SVD modes of the reconstructed state $X_K$ across temperatures: the slow eigenfunction mainly splits the state space along the position coordinate and the additional high velocity $|u_2|$ transition regions.
}
\label{fig_S_DW_phi2}
\end{center}
\end{figure*}

\begin{figure*}
\begin{center}
\includegraphics[scale=1]{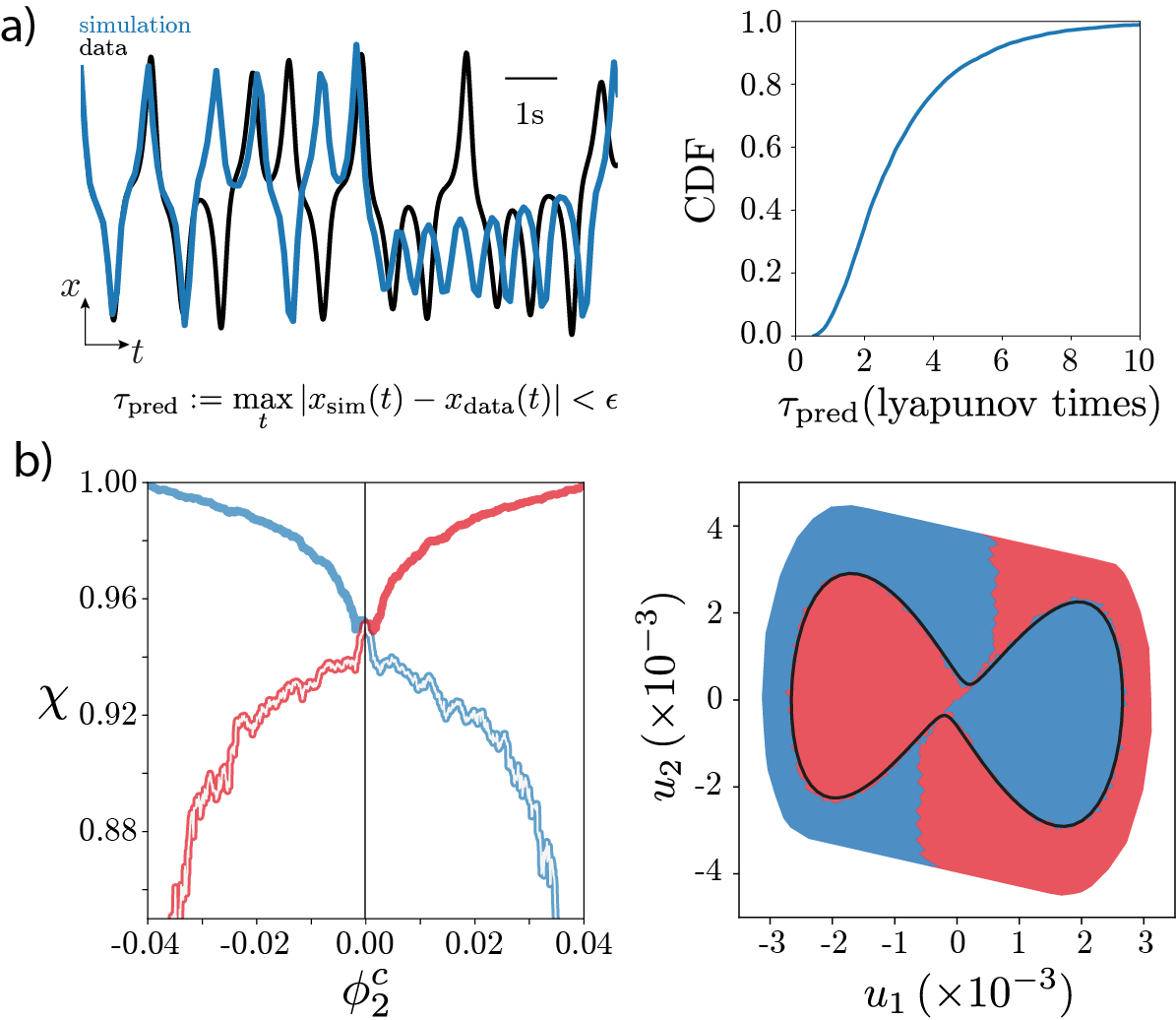}
\caption{{\bf Coarse-graining in the Lorenz system.}
(a) We assess the predictive ability of our inferred reconstructed transfer operator by simulating time series of the $x$ measurements as in Fig.\,\ref{fig_2}(b-bottom). Starting from an initial partition $s_i$, we sample from the corresponding row of the inferred Markov chain $P_{ij}(\tau^*)$ generating a symbolic sequence sampled at $\delta t = \tau^*$. From each symbol we then sample a state space point at random within the corresponding partition, from which we extract $x$ (see Appendix \ref{sec:Methods}). We use $K^*=12\,\text{frames}$, $N^*=10^{3.5}\,\text{partitions}$ and $\tau^* = 0.1\,\text{s}$, Figs.\,\ref{fig_2}(b),\ref{fig_3_Lorenz}(a). To quantitatively assess the predictive power of this simple model, we define the prediction time $\tau_\text{pred}$ as the time it takes for two trajectories, $x_\text{sim}$ and $x_\text{data}$ to deviate more that $\epsilon = \frac{\max(x) - \min(x)}{2}$, which essentially capture the hopping between lobes of the attractor. We start $x_\text{sim}$ and $x_\text{data}$ from the same partition, but evolve them either according to the $P_{ij}(\tau^*)$ or Eq.\,\ref{eq:Lorenz}, respectively. We find that our model is predictive on relatively long time scales, with $\langle \tau_\text{pred} \rangle = 3.04\,(2.99,3.08)\,\text{lyapunov times}$.
(b-left) Degree of almost invariance as measure by $\chi$, Eq.(\ref{Eq:coherence_min}). Colors represent the contribution of each almost-invariant set ($\chi_{\mu,\tau}(S^+)$ and $\chi_{\mu,\tau}(S^-)$) while the white line is the minimum between sets, $\chi$. We split the state space along the global maximum of $\chi$, $\phi_2^c\sim 0$.
(b-right) We visualize the almost-invariant sets obtained by partitioning the state space along $\phi_2^c$ by projecting onto the first two singular vectors of the trajectory matrix, $X_K^*$. The boundary between almost-invariant sets is partially defined by the stable manifold of the shortest period UPO \cite{Froyland2009}, of which we show only the orbit itself (identified through recurrences) for clarity (Appendix \ref{sec:Methods}).
}
\label{fig_SLorenz_AIS}
\end{center}
\end{figure*}

\begin{figure*}
\begin{center}
\includegraphics[scale=1]{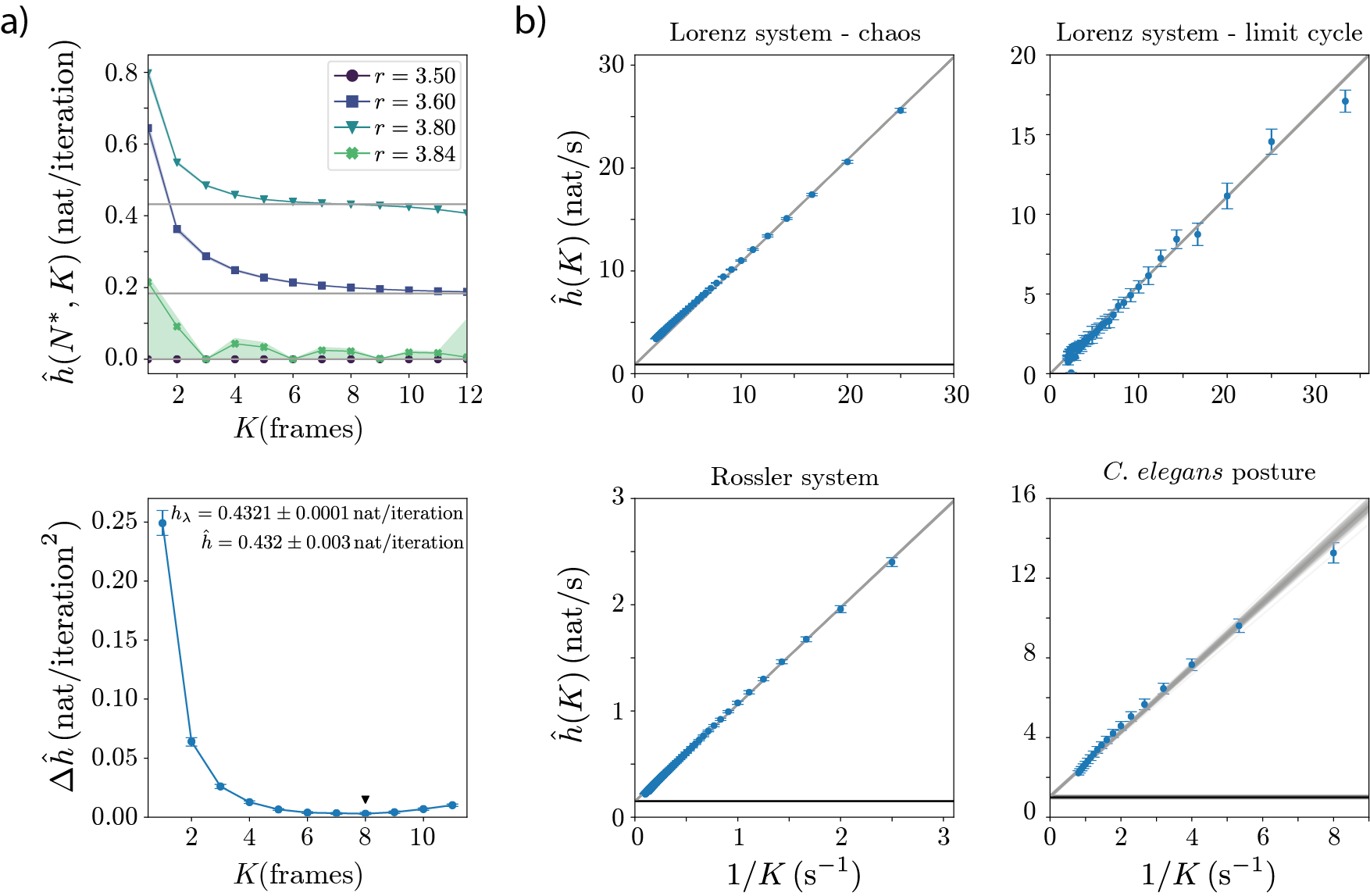}
\caption{{\bf KS entropy estimation in discrete maps and continuous-time dynamical systems.}
(a-top) Entropy rate estimates $\hat{h}(N^*,K)$ in the Logistic map, Eq.\,\ref{eq:Logistc}, for different representative values of the control parameter $r$, with $N^*$ chosen so as to maximize $\hat{h}(N^*,K)$ for each $K$. The horizontal lines represent the positive Lyapunov exponents estimated from the equations of motion (shaded regions represent 95\% confidence intervals bootstrapped across 100 segments with $T=10^7$ iterations). 
(a-bottom) We find $K^*$ by examining the local slope of the entropy rate, which becomes higher when finite-size effects are present: our best estimate of the entropy rate is $\hat{h}(K^*)$ corresponding to $\min_{K^*}\Delta\hat{h}(K)$, Eq.\,\ref{eq:delta_h}, yielding an accurate estimate of the positive Lyapunov exponent $h_\lambda$ across $r$, Fig.\,\ref{fig4}(a).
(b) Unlike the simple discrete maps, we find that for continuous-time dynamics with short $\delta t$ the value of $K$ needed to accurately approximate the KS entropy is beyond what can be directly sampled. To overcome this difficulty, we estimate the $K\rightarrow\infty$ of the entropy rate limit through extrapolation as in Ref.~\onlinecite{Strong1998}. For increasing $K$, we observe that the maximum entropy  $\hat{h}(K,N)$ with respect to N, $\hat{h}(K) = \max_N \hat{h}(K,N)$ behaves as $1/K$ for a wide range of $K$ values, suggesting that we can estimate the $K\rightarrow\infty$ limit of the KS entropy through extrapolation.
We show the result of weighted least squares regression (gray lines) of $\hat{h} = m/K+h_\infty$. The estimate of the KS entropy through the sum of Lyapunov exponents is shown as a black horizontal line; for the toy systems we used the algorithm of Ref.~\onlinecite{Chen2006}, while for the \emph{C. elegans} data we show the results of Ref.~\onlinecite{Ahamed2021}.
}
\label{figS_KS_entropy}
\end{center}
\end{figure*}

\begin{figure*}
\begin{center}
\includegraphics[scale=1]{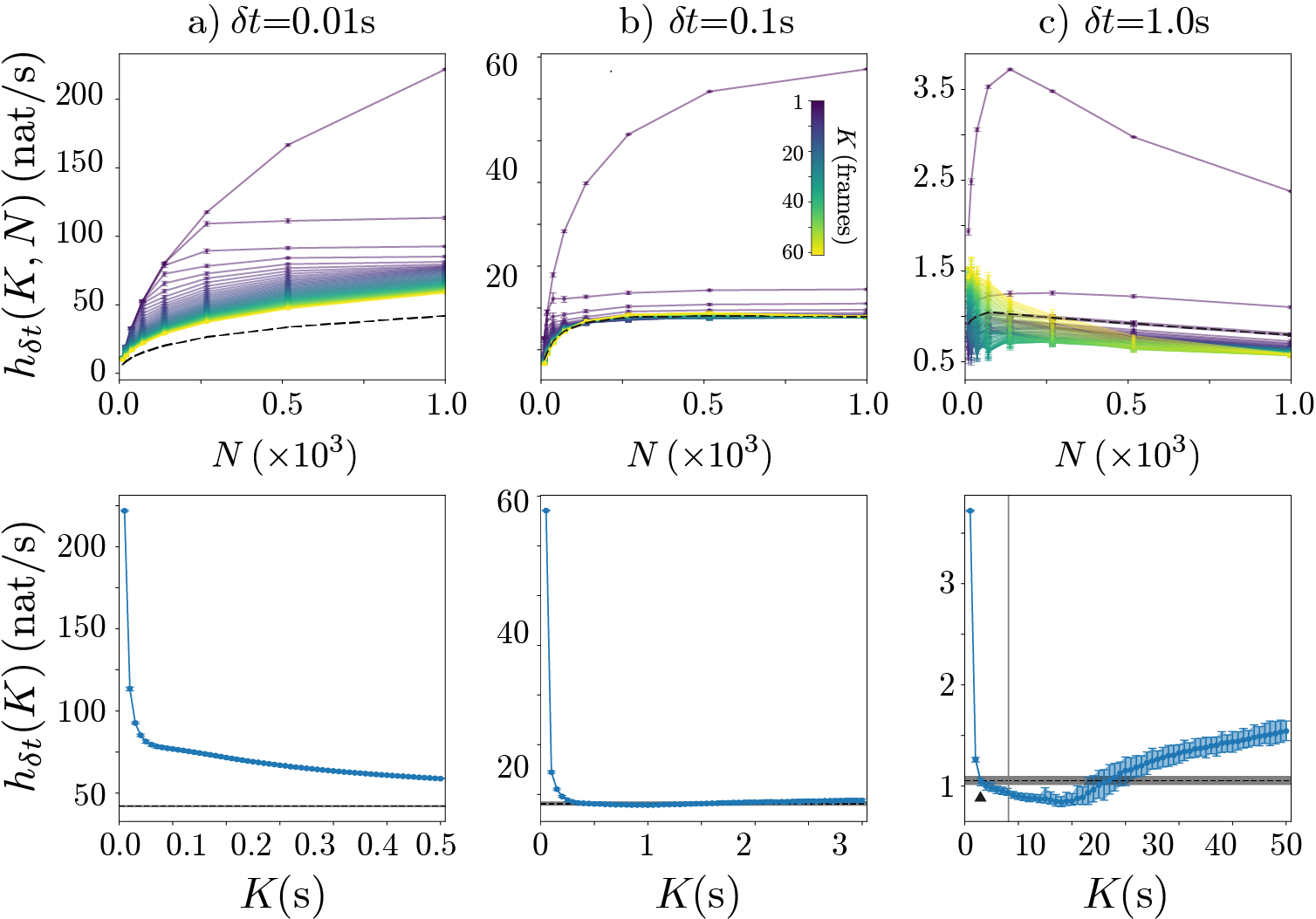}
\caption{{\bf Behavior of the short-time entropy rate $h_{\delta t}(N,K)$ with sampling time $\delta t$ in the R\"{o}ssler system.} We estimate the short-time entropy rate, Eq.\,\ref{eq:h_p}, of the reconstructed state space obtained from measuring the $x$ variable of the R\"{o}ssler system. We plot $h_{\delta t}(K,N)$ as a function of $N$ (top), for increasing number of time delays $K$ (color coded) and the corresponding $h_{\delta t}(K) = \max_N h_{\delta t}(K,N)$ (bottom) for different values of $\delta t$. The black dashed lines represents the short-time entropy rate computed from the original 3-dimensional state space, Eq.\,\ref{eq:Rossler}. 
When $\delta t$ is too short (a), the $K$ required to reconstruct the state space is large, making it computationally impractical to achieve. In contrast, when $\delta t$ is long (c), we reach the short-time entropy rate of the underlying dynamics after only $K=3\,\text{frames}$, which corresponds to the dimensionality of the state space. Notably, beyond this point the estimates of the entropy rate deviate from that of the underlying state space. For $5\lesssim K \lesssim 25$, we observe an underestimation of the entropy-rate likely due to finite-size effects. On the other hand, for $K\gtrsim 30$ the sequences $x_{t:t+K}$ used to compute the entropy rate are uncorrelated vectors that span the entire state space (the slowest time scale of the dynamics as estimated through the power spectral density is $\tau_\text{slow} = 8.17\,(8.02, 8.30)\,\text{s}$, indicated with a gray vertical line on c-bottom). This means that the state-space reconstruction is composed of independent $K$-dimensional vectors constructed from points sampled uniformly on the attractor: with the increasing dimensionality $K$ of this unstructured space, the k-means partitioning becomes plagued by the curse of dimensionality resulting in an increase of $h_{\delta t}(N^*,K)$ for large $K$. For $\delta t= 0.1\,\text{s}$ (b) the finite size effects become negligible: $\delta t$ is short enough to capture the fast time scales of the dynamics, while being long enough so as not too yield a state space reconstruction that is too high-dimensional as in (a).
}
\label{fig_S_dt}
\end{center}
\end{figure*}

\end{document}